\documentclass[aip,amsmath,amssymb,reprint]{revtex4-1}

\usepackage{geometry}
\usepackage[T1]{fontenc}

\usepackage{amsmath,color}
\usepackage{amssymb}
\usepackage{amsfonts}
\usepackage{amsthm}
\usepackage{mathtools}
\usepackage{physics}
\usepackage{multirow}
\usepackage{comment}
\usepackage{xcolor} 
\usepackage{bm}

\usepackage[colorlinks=true, linkcolor=black, citecolor=black, filecolor=blue, urlcolor=blue]{hyperref}
\hypersetup{
    linkcolor=black,
    citecolor=black,
    filecolor=black,
    urlcolor=blue,
    linkbordercolor=white,
    citebordercolor=white,
    urlbordercolor=white,
    pdfborder={0 0 0}
}

\usepackage{threeparttable}
\usepackage{placeins}

\usepackage[labelfont=bf]{caption}
\usepackage{ragged2e} 

\DeclareCaptionLabelSeparator{boldperiod}{\textbf{.\ }}
\renewcommand{\thefigure}{\arabic{figure}} 
\renewcommand{\thetable}{\Roman{table}} 

\DeclareCaptionType{scheme}[SCHEME][List of Schemes]
\renewcommand{\thescheme}{\Roman{scheme}}

\usepackage{algorithm}
\usepackage{algpseudocode}
\usepackage{dsfont}

\usepackage{accents}
\definecolor{active_color}{named}{black} 

\usepackage{lipsum}

\newcommand{\mr}[1]{\mathrm{#1}}
\newcommand{\mb}[1]{\mathbf{#1}}
\newcommand{\mc}[1]{\mathcal{#1}}

\begin{document}

\title[]{\Large Extended Lagrangian molecular dynamics on vibronic surfaces in the nuclear--electronic orbital framework}

\author{Joseph A. Dickinson}
\affiliation{\small Department of Chemistry, Yale University, New Haven, CT 06520, USA}
\affiliation{\small Department of Chemistry, Princeton University, Princeton, NJ 08544, USA}

\author{Mathew Chow}
\affiliation{\small Department of Chemistry, Yale University, New Haven, CT 06520, USA}
\affiliation{\small Department of Chemistry, Princeton University, Princeton, NJ 08544, USA}

\author{Eno Paenurk}
\affiliation{\small Department of Chemistry, Princeton University, Princeton, NJ 08544, USA}

\author{Sharon Hammes-Schiffer}
\email{shs566@princeton.edu}
\affiliation{\small Department of Chemistry, Yale University, New Haven, CT 06520, USA}
\affiliation{\small Department of Chemistry, Princeton University, Princeton, NJ 08544, USA}

\begin{abstract}
    \noindent Proton transfer is central to many processes of chemical interest. The simulation of proton transfer dynamics requires the inclusion of nuclear quantum effects, such as zero-point energy, nuclear delocalization, and tunneling.  Herein, we introduce methods within the nuclear--electronic orbital (NEO) framework, where specified nuclei are treated quantum mechanically on the same level as the electrons, for the simulation of proton transfer dynamics. Specifically, NEO density functional theory is used to treat the transferring protons quantum mechanically, and the other nuclei are propagated classically on the adiabatic vibronic ground-state surface. We formulate a NEO extended Lagrangian molecular dynamics (NEO-ELMD) approach to incorporate the motion of the nuclear basis function centers during such simulations. Density matrix extrapolation and purification are introduced as a means to accelerate the NEO self-consistent field procedure at each time step by reducing the number of iterations required for convergence. We demonstrate the fidelity and efficiency of NEO-ELMD by comparison to related dynamics methods for intramolecular proton transfer in malonaldehyde. We also use these accelerated techniques to simulate the nonequilibrium single and double proton transfer dynamics of proton-coupled electron transfer in much larger benzimidazole-phenol systems. This work provides a foundation for future methodologies to efficiently simulate proton transfer dynamics within the NEO-DFT framework while incorporating nonadiabatic effects between adiabatic vibronic states.
\end{abstract}

\maketitle



\section{Introduction}\label{sec:intro}

Proton transfer plays a crucial role in a wide range of chemical and biological processes. \cite{stubbe_radical_2003, hammes-schiffer_hydrogen_2006, hammarstrom_proton-coupled_2011} Nuclear quantum effects (NQEs), such as nuclear delocalization and zero-point energy, often significantly impact proton transfer dynamics.\cite{markland_nuclear_2018} The significance of NQEs has motivated the development of a large body of methods to capture them during dynamics simulations. Wavepacket dynamics \cite{meyer_multi-configurational_1990, beck_multiconfiguration_2000, meyer_quantum_2003} provide accurate results but generally cannot be scaled to large molecular or condensed-phase systems due to computational expense. Computational efficiency can be gained by employing moving nuclear basis sets, as in  thawed\cite{heller_timedependent_1975} and frozen \cite{heller_frozen_1981} Gaussian approaches for wave packet simulations, direct dynamics with variational Gaussian wave packets, \cite{worth_novel_2004} and the multiple spawning method. \cite{ben-nun_ab_2000, ben-nun_ab_2002} Approaches based on quantum mechanical path integrals, such as the semiclassical initial value representation (SC-IVR), \cite{wang_semiclassical_1998, sun_semiclassical_1998} path integral molecular dynamics (PIMD), \cite{berne_classical_1998, tuckerman_path_2002} and ring-polymer molecular dynamics (RPMD), \cite{craig_quantum_2004, habershon_ring-polymer_2013} capture NQEs within trajectory-based simulations. However, path integral methods rely on the propagation of the coupled dynamics of many replicas of the system. 

Multicomponent quantum chemistry approaches \cite{kreibich_multicomponent_2001, pavosevic_multicomponent_2020} provide a promising avenue for capturing NQEs on-the-fly during first-principles molecular dynamics simulations. In particular, the nuclear–electronic orbital (NEO) \cite{webb_multiconfigurational_2002} approach has emerged as a prominent framework for capturing NQEs in this manner. In the NEO approach, select nuclei are quantized and treated on an equal footing as the electrons within quantum chemistry calculations. \cite{hammes-schiffer_nuclearelectronic_2021} The mixed nuclear--electronic Schr\"odinger equation is solved in a self-consistent manner, providing  both electronic and nuclear densities while incorporating the NQEs directly into the NEO energies. In addition, the NEO approach avoids the Born--Oppenheimer separation between the quantum nuclei and the electrons.

A large suite of NEO methods has been developed, \cite{webb_multiconfigurational_2002, pavosevic_multicomponent_2019, pavosevic_multicomponent_2020, hammes-schiffer_nuclearelectronic_2021, alaal_multicomponent_2021, hasecke_local_2024, malbon_nuclearelectronic_2025, goudy_triple_2025} with NEO density functional theory (NEO-DFT) \cite{pak_density_2007, brorsen_multicomponent_2017, yang_development_2017, yang_multicomponent_2018} emerging as the main methodology for simulating  proton transfer dynamics, \cite{tao_direct_2021, zhao_real-time_2020, zhao_excited_2021, li_semiclassical_2022, li_nuclear-electronic_2023, chow_nuclearelectronic_2023, chow_nuclearelectronic_2023-1, chow_nuclear_2024} although NEO time-dependent configuration interaction \cite{garner_nuclearelectronic_2024, garner_time-resolved_2025} is another promising direction. The popularity of NEO-DFT for dynamical simulations is due to the efficient inclusion of electron--proton correlation (epc) via the epc functionals.\cite{brorsen_multicomponent_2017, yang_development_2017} In this work, we will assume that the underlying nuclear--electronic structure of all systems is solved with NEO-DFT and that all quantized nuclei are protons. We will also limit our discussion to dynamics on the electron--proton vibronic ground-state surface, with the understanding that the methods discussed herein can be combined with nonadiabatic dynamics approaches, \cite{tully_perspective_2012} such as Ehrenfest \cite{li_ab_2005, tully_ehrenfest_2023} and surface hopping \cite{tully_molecular_1990, hammesschiffer_proton_1994} methods, to describe dynamics on multiple vibronic surfaces.  

The dynamics of the classical nuclei on the electron--proton vibronic ground-state surface can be simulated with NEO Born--Oppenheimer molecular dynamics (NEO-BOMD), as demonstrated for hydride transfer in C$_4$H$_9^+$. \cite{schneider_transition_2021, tao_direct_2021} Analogous to conventional BOMD, \cite{marx_ab_2009} the NEO-DFT ground state energy is computed via the NEO self-consistent field (NEO-SCF) procedure at each time step of a molecular dynamics (MD) trajectory. The gradient of this energy is then used to numerically propagate the classical nuclei according to Newtonian equations of motion, solving the NEO-SCF equations at each time step. The protonic basis function center positions can be optimized variationally at each time step to obtain the lowest-energy vibronic state energy. 

NEO-BOMD simulations are typically more expensive than their conventional BOMD counterparts due in part to the slightly higher cost of the NEO-SCF procedure. However, the speed of the NEO-SCF procedure can be enhanced via simultaneous optimization of nuclear and electronic orbitals,\cite{liu_simultaneous_2022, chow_efficient_2026} the nuclear Hartree product representation for systems with multiple quantum protons, \cite{auer_localized_2010, chow_efficient_2026} and more sophisticated initial guesses. \cite{chow_efficient_2026} Although these enhancements do indeed offer a substantial speedup, obtaining long-time nuclear dynamics, on the order of picoseconds, is still a challenge. Similarly, conventional BOMD simulations have been made more efficient via techniques such as extrapolation procedures for Fock \cite{pulay_fock_2004} or density \cite{kuhne_efficient_2007, polack_grassmann_2021} matrix elements, yet  long-time dynamics is still challenging.

Extended Lagrangian approaches \cite{car_unified_1985, head-gordon_curvy_2003, niklasson_extended_2008, niklasson_extended_2009} have also been used to enhance the efficiency of conventional BOMD simulations, although they do not yield exact classical dynamics on Born--Oppenheimer potential energy surfaces. In extended Lagrangian approaches for BOMD, fictitious degrees of freedom are introduced into the system and are governed by underlying equations of motion in an effort to circumvent costly computations. The most famous extended Lagrangian approach to BOMD is the Car--Parrinello molecular dynamics approach.\cite{car_unified_1985} Although the classical nuclei do not evolve on the exact potential energy surface for these approaches, the dynamics are qualitatively correct for sufficiently small fictitious masses.

In NEO simulations, an extended Lagrangian approach \cite{xu_lagrangian_2024} can be employed to move the nuclear basis functions during dynamics via a traveling proton basis (TPB) scheme. Until now, TPB schemes have only been applied to real-time NEO dynamics methods,\cite{zhao_real-time_2020, zhao_nuclear-electronic_2020, zhao_excited_2021, xu_lagrangian_2024, li_energy_2025} which propagate the nuclear-electronic time-dependent Schr\"odinger equation. In these TPB schemes, the nuclear basis function centers are propagated classically along with the classical nuclei, in conjunction with the time-dependent quantum dynamical propagation of the electronic and protonic densities. These real-time NEO dynamics methods are distinct from NEO-BOMD on the adiabatic vibronic ground state, where the nuclear--electronic time-independent Schr\"odinger equation is solved via the NEO-SCF procedure at each MD time step. 

Herein, we introduce a NEO extended Lagrangian MD (NEO-ELMD) method to perform NEO-BOMD at a greatly reduced cost. Adopting the always stable predictor-corrector scheme, \cite{kolafa_numerical_1996, kolafa_time-reversible_2004, kuhne_efficient_2007} we show that density matrix extrapolation and subsequent purification \cite{mcweeny_recent_1960} can enhance the speed of NEO-SCF procedures performed at each time step of a NEO-ELMD trajectory. We also show that an extended Lagrangian approach for moving nuclear basis function centers during such simulations yields qualitatively similar results to NEO-BOMD with variationally optimized nuclear basis function centers. In addition, we compare these NEO-ELMD approaches to constrained NEO molecular dynamics (CNEO-MD), which applies a Lagrangian constraint such that the nuclear basis function center remains at the expectation value of the nuclear position operator. \cite{xu_constrained_2020, xu_molecular_2022, chen_incorporating_2023} We discuss how CNEO-MD arises as a special case of NEO-ELMD and show that the efficiency of CNEO-MD is also enhanced by the density matrix extrapolation procedures. We benchmark the new methods by simulating the intramolecular proton transfer of malonaldehyde. We then use these NEO-ELMD methods to simulate the nonequilibrium dynamics of proton-coupled electron transfer (PCET) in much larger benzimidazole-phenol (BIP) constructs. \cite{odella_controlling_2018, odella_proton-coupled_2019, odella_proton-coupled_2020, goings_nonequilibrium_2020, yoneda_electronnuclear_2021, arsenault_concerted_2022}

This paper is organized as follows. Section~\ref{sec:theory} provides an overview of the NEO-ELMD method, including the details of the density matrix extrapolation procedures and TPB scheme. Section~\ref{sec:comp_details} provides the computational details of the malonaldehyde and BIP trajectories, and Section~\ref{sec:results} provides the results of these simulations. We discuss future directions in Section~\ref{sec:conclusions}.

\section{Theory and Methods}\label{sec:theory}

\subsection{Overview of NEO-DFT}\label{subsec:neo_dft_overview}

Consider a system composed of an $N^{\mr{e}}$ electrons,  $N^{\mr{p}}$ quantum protons, and $N^{\mr{c}}$ classical nuclei. The non-relativistic Hamiltonian for this system, $\hat{H}_{\mr{NEO}}$, is given in atomic units by
\begin{equation}\label{eq:H_NEO}
\begin{aligned}
    \hat{H}_{\mr{NEO}} &= -\frac{1}{2}\sum_i^{N^{\mr{e}}} \nabla_i^2 - \frac{1}{2m_{\mr{p}}}\sum_I^{N^{\mr{p}}} \nabla_I^2 \\
    &+ \sum_i^{N^{\mr{e}}}\sum_{j>i}^{N^{\mr{e}}} \frac{1}{\lvert\mb{r}_i -\mb{r}_j\rvert} + \sum_I^{N^{\mr{p}}}\sum_{J>I}^{N^{\mr{p}}} \frac{1}{\lvert\mb{r}_I -\mb{r}_J\rvert}  \\
    &- \sum_i^{N^{\mr{e}}}\sum_A^{N^{\mr{c}}} \frac{Z_{\mr{A}}}{\lvert\mb{r}_i-\mb{R}_A\rvert} + \sum_I^{N^{\mr{p}}}\sum_A^{N^{\mr{c}}} \frac{Z_{\mr{A}}}{\lvert\mb{r}_I-\mb{R}_A\rvert} \\
    &- \sum_i^{N^{\mr{e}}}\sum_I^{N^{\mr{p}}}  \frac{1}{\lvert\mb{r}_i -\mb{r}_I\rvert} + \sum_A^{N^{\mr{c}}}\sum_{B>A}^{N^{\mr{c}}} \frac{Z_{\mr{A}} Z_{\mr{B}}}{\lvert\mb{R}_A-\mb{R}_B\rvert},
\end{aligned}
\end{equation}
where the indices $\{i,j,\dots\}$, $\{I,J,\dots\}$, and $\{A,B,\dots\}$ denote electrons, quantum protons, and classical nuclei, respectively, $m_\mr{p}$ is the mass of a proton, $\mb{r}$ denotes the position of either an electron or a quantum proton, and $\mb{R}$ denotes the position of a classical nucleus. 

Within multicomponent DFT, the energy of this system, $U_{\mr{NEO}}$, is a functional of the electronic and protonic densities, $\rho^{\mr{e}}$ and $\rho^{\mr{p}}$, respectively, according to
\begin{equation}
\begin{aligned}\label{eq:U_NEO}
    U_{\mr{NEO}}[\rho^{\mr{e}}, \rho^{\mr{p}}] &=  U_{\mr{ref}}[\rho^{\mr{e}}, \rho^{\mr{p}}] + U_{\mr{ext}}[\rho^{\mr{e}}, \rho^{\mr{p}}] \\
    &+ U_{\mr{exc}}[\rho^{\mr{e}}] + U_{\mr{pxc}}[\rho^{\mr{p}}] + U_{\mr{epc}}[\rho^{\mr{e}}, \rho^{\mr{p}}],
\end{aligned}
\end{equation}
Here, $U_{\mr{ref}}[\rho^{\mr{e}}, \rho^{\mr{p}}]$ is the energy of the noninteracting reference system of electrons and quantum protons, including the kinetic energies of the electrons and quantum protons and the Coulomb interactions within this quantum subsystem. $U_{\mr{ext}}[\rho^{\mr{e}}, \rho^{\mr{p}}]$ corresponds to the interaction of the electronic and protonic densities with the external field generated by the  classical nuclei. $U_{\mr{exc}}[\rho^{\mr{e}}]$, $U_{\mr{pxc}}[\rho^{\mr{p}}]$, and $U_{\mr{epc}}[\rho^{\mr{e}}, \rho^{\mr{p}}]$ denote the electron-electron exchange-correlation functional, proton-proton exchange-correlation functional, and electron-proton correlation functional, respectively. 

In NEO-DFT, the KS reference state is represented by a product of an electronic and protonic Slater determinant. \cite{pavosevic_multicomponent_2020} However, due to the locality of quantized protons in molecular systems, the nuclear Slater determinant can be replaced by the nuclear Hartree product representation, treating the quantum protons as distinguishable. \cite{auer_localized_2010} This treatment is justified because the proton-proton exchange terms are typically eight to ten orders of magnitude smaller than their electronic counterparts. \cite{pavosevic_multicomponent_2020, goudy_triple_2025} The nuclear Hartree product representation greatly reduces the computational cost of the NEO-SCF procedure, \cite{auer_localized_2010} mainly by significantly decreasing the number of iterations compared to the nuclear Slater determinant representation. \cite{chow_efficient_2026} Thus, in this work, all calculations with multiple quantum protons use the nuclear Hartree product representation.

In this context, the NEO-DFT KS reference state $\ket{\Psi(\mb{x}^{\mr{e}},\mb{x}^{\mr{p}})}$ is expressed as
\begin{equation}
    \ket{\Psi(\mb{x}^{\mr{e}},\mb{x}^{\mr{p}})} = \Phi^{\mr{e}}(\mb{x}^{\mr{e}})\prod_{I=1}^{N^{\mr{p}}} \chi_I^{\mr{p}}(\mb{x}_I^{\mr{p}}),
\end{equation}
where $\mb{x}^{\mr{e(p)}}$ are generalized electronic (protonic) coordinates containing position and spin, $\chi_I^{\mr{p}}$ is the spin orbital for the $I$-th quantum proton, and $\Phi^{\mr{e}}$ is an electronic Slater determinant. Assuming a closed-shell electronic configuration, the electronic and protonic densities $\rho^{\mr{e}}$ and $\rho^{\mr{p}}$ can be expressed in terms of doubly-occupied electronic spatial orbitals $\{\psi_{i}^{\mr{e}}\}$ and singly-occupied protonic spatial orbitals $\{\psi_{I}^{\mr{p}}\}$ as
\begin{subequations}\label{eq:rho}
\begin{align}
    \rho^{\mr{e}}(\mb{r}_1^{\mr{e}}) &= 2\sum_{i=1}^{N^{\mr{e}}/2} \lvert\psi_i^{\mr{e}}(\mb{r}_1^{\mr{e}})\rvert^2, \label{eq:rho-e} \\
    \rho^{\mr{p}}(\mb{r}_1^{\mr{p}}) &= \sum_{I=1}^{N^{\mr{p}}} \lvert\psi_I^{\mr{p}}(\mb{r}_1^{\mr{p}})\rvert^2  = \sum_{I=1}^{N^{\mr{p}}} \rho^{\mr{p}}_I(\mb{r}_1^{\mr{p}}), \label{eq:rho-p}
\end{align}
\end{subequations}
where the density of the $I$-th distinguishable quantum proton is defined as $\rho^{\mr{p}}_I = \lvert\psi_I^{\mr{p}}\rvert^2$. Generalization to open-shell electronic configurations is straightforward. 

Minimizing the NEO-DFT energy given in Eq.~\eqref{eq:U_NEO} with respect to electronic and protonic orbital variations under orthonormality constraints leads to the NEO-KS equations, 
\begin{subequations}\label{eq:neo_ks}
\begin{align}
    \left(-\frac{1}{2}\nabla_1^2 + v_{\mr{eff}}^{\mr{e}}(\mb{r}_1^{\mr{e}})\right)\psi_i^{\mr{e}}(\mb{r}_1^{\mr{e}}) &= \epsilon_i^{\mr{e}}\psi_i^{\mr{e}}(\mb{r}_1^{\mr{e}}), \label{eq:neo_ks-e} \\
    \left(-\frac{1}{2m_{\mr{p}}}\nabla_1^2 + v_{\mr{eff}, I}^{\mr{p}}(\mb{r}_1^{\mr{p}})\right)\psi_I^{\mr{p}}(\mb{r}_1^{\mr{p}}) &= \epsilon_I^{\mr{p}}\psi_I^{\mr{p}}(\mb{r}_1^{\mr{p}}), \label{eq:neo_ks-p}
\end{align}
\end{subequations}
where 
\begin{widetext}
\begin{subequations}\label{eq:veff}
\begin{align}
    v_{\mr{eff}}^{\mr{e}}(\mb{r}_1^{\mr{e}}) &=  -\sum_A^{N^{\mr{c}}} \frac{Z_{\mr{A}}} {\lvert\mb{r}_1^{\mr{e}}-\mb{R}_A\rvert} + \int d\mb{r}_2^{\mr{e}}\frac{\rho^{\mr{e}}(\mb{r}_2^{\mr{e}})}{\lvert\mb{r}_1^{\mr{e}}-\mb{r}_2^{\mr{e}}\rvert} - \int d\mb{r}_1^{\mr{p}}\frac{\rho^{\mr{p}}(\mb{r}_1^{\mr{p}})}{\lvert\mb{r}_1^{\mr{e}}-\mb{r}_1^{\mr{p}}\rvert} + \frac{\delta U_{\mr{exc}}[\rho^{\mr{e}}]}{\delta\rho^{\mr{e}}(\mb{r}_1^{\mr{e}})} + \frac{\delta U_{\mr{epc}}[\rho^{\mr{e}},\rho^{\mr{p}}]}{\delta\rho^{\mr{e}}(\mb{r}_1^{\mr{e}})}, \label{eq:veff-e} \\
    v_{\mr{eff}, I}^{\mr{p}}(\mb{r}_1^{\mr{p}}) &= \sum_A^{N^{\mr{c}}} \frac{Z_{\mr{A}}}{\lvert\mb{r}_1^{\mr{p}}-\mb{R}_A\rvert} + \sum_{J\neq I}^{N^{\mr{p}}}\int d\mb{r}_2^{\mr{p}}\frac{\rho_J^{\mr{p}}(\mb{r}_2^{\mr{p}})}{\lvert\mb{r}_1^{\mr{p}}-\mb{r}_2^{\mr{p}}\rvert} - \int d\mb{r}_1^{\mr{e}}\frac{\rho^{\mr{e}}(\mb{r}_1^{\mr{e}})}{\lvert\mb{r}_1^{\mr{e}}-\mb{r}_1^{\mr{p}}\rvert} + \frac{\delta U_{\mr{epc}}[\rho^{\mr{e}},\rho^{\mr{p}}]}{\delta\rho^{\mr{p}}(\mb{r}_1^{\mr{p}})},
\end{align}
\end{subequations}
\end{widetext}
and $\epsilon_{i(I)}^{\mr{e(p)}}$ is the orbital energy of the corresponding electronic (protonic) orbital. This expression neglects the proton-proton correlation energy, which is valid due to the locality of the quantum protons as well as previous work  showing that proton-proton correlation terms are many orders of magnitude smaller than their electronic counterparts.\cite{pavosevic_multicomponent_2020, goudy_triple_2025}

The electronic (protonic) orbitals can be expanded in terms of electronic (protonic) basis functions $\{\phi_{\mu}^{\mr{e(p)}}\}$ according to 
\begin{equation}
    \psi_{i(I)}^{\mr{e(p)}} = \sum_\mu^{N_{\mr{basis}}^{\mr{e(p)}}} C_{\mu, i(I)}\phi_\mu^{\mr{e(p)}},
\end{equation}
where $N_{\mr{basis}}^{\mr{e(p)}}$ is the number of electronic (protonic) basis functions. This expansion leads to a coupled set of NEO-KS matrix equations
\begin{subequations}\label{eq:roothan}
\begin{align}
    \mb{F}^{\mr{e}} \mb{C}^{\mr{e}} &= \mb{S}^{\mr{e}}\mb{C}^{\mr{e}}\bm{\epsilon}^{\mr{e}}, \label{eq:roothan-e} \\
    \mb{F}^{\mr{p}} \mb{C}^{\mr{p}} &= \mb{S}^{\mr{p}}\mb{C}^{\mr{p}}\bm{\epsilon}^{\mr{p}}, \label{eq:roothan-p}
\end{align}
\end{subequations}
where $\mb{F}^{\mr{e(p)}}$, $\mb{C}^{\mr{e(p)}}$, $\mb{S}^{\mr{e(p)}}$, and $\bm{\epsilon}^{\mr{e(p)}}$ are the electronic (protonic) Kohn-Sham, coefficient, overlap, and orbital energy matrices, respectively. These equations are strongly coupled because both the electronic and protonic Fock matrices depend on both the electronic and protonic coefficient matrices. Eq.~\eqref{eq:roothan} is solved self-consistently via the NEO-SCF procedure,\cite{liu_simultaneous_2022, chow_efficient_2026} yielding converged electronic and protonic densities, as well as the NEO-DFT energy given in Eq.~\eqref{eq:U_NEO}. The analytical gradient of this NEO-DFT energy is given in Eq. (17) of Ref. ~\citenum{tao_direct_2021}.

In typical NEO calculations, each quantum proton is associated with a single protonic basis function center. A constraint can be imposed to ensure that the expectation value of each quantum proton position operator coincides with its respective protonic basis function center position, as in constrained NEO (CNEO) DFT. \cite{xu_constrained_2020} In the CNEO-DFT approach, $U_{\mr{NEO}}$ in Eq.~\eqref{eq:U_NEO} is minimized subject to orthonormality constraints on the electronic and protonic orbitals, as well as $N^{\mr{p}}$ constraints on the protonic orbitals
\begin{equation}\label{eq:cneo_constraints}
    \mel{\psi_I^{\mr{p}}}{\hat{\mb{r}}}{\psi_I^{\mr{p}}} = \mc{R}_I, \hspace{0.5em}\forall I\in\{1,\dots,N^{\mr{p}}\},
\end{equation}
where $\mc{R}_I$ is the position of the protonic basis function center associated with the $I$-th quantum proton. Imposing these additional constraints leads to slightly modified NEO-KS equations, 
\begin{subequations}\label{eq:cneo_ks}
\begin{align}
    \left(-\frac{1}{2}\nabla_1^2 + v_{\mr{eff}}^{\mr{e}}(\mb{r}_1^{\mr{e}})\right)\psi_i^{\mr{e}}(\mb{r}_1^{\mr{e}}) &= \epsilon_i^{\mr{e}}\psi_i^{\mr{e}}(\mb{r}_1^{\mr{e}}), \label{eq:cneo_ks-e} \\
    \left(-\frac{1}{2m_{\mr{p}}}\nabla_1^2 + v_{\mr{eff}, I}^{\mr{p}}(\mb{r}_1^{\mr{p}})+\mb{f}_I\cdot\mb{r}_1^{\mr{p}}\right)\psi_I^{\mr{p}}(\mb{r}_1^{\mr{p}}) &= \epsilon_I^{\mr{p}}\psi_I^{\mr{p}}(\mb{r}_1^{\mr{p}}), \label{eq:cneo_ks-p}
\end{align}
\end{subequations}
where $\mb{f}_I$ is a three-component vector of Lagrange multipliers for the $I$-th proton that satisfies the $I$-th constraint of Eq.~\eqref{eq:cneo_constraints}. Note that because the constraints of Eq.~\eqref{eq:cneo_constraints} only act on the protonic orbitals, only the eigenvalue equations for the quantum protons are changed in Eq.~\eqref{eq:cneo_ks} compared to  Eq.~\eqref{eq:neo_ks}.  The coupled NEO-KS matrix equations of Eq.~\eqref{eq:roothan} are now solved via the CNEO-SCF routine, which includes optimization of all $N^{\mr{p}}$ Lagrange multiplier vectors $\mb{f}_I$ needed to satisfy each constraint of Eq.~\eqref{eq:cneo_constraints}. We refer the reader to Eq. (S19) of Ref.~\citenum{xu_full-quantum_2020} for the analytical gradient of $U_{\mr{NEO}}$ when subject to the constraints of Eq.~\eqref{eq:cneo_constraints}. It can be shown that the optimized $\mb{f}_I$ for each quantum proton is equal to the force on the expectation value of that proton to constrain it to that position. \cite{xu_constrained_2020}

\subsection{NEO Extended Lagrangian Molecular Dynamics}\label{subsec:elmd}

In conventional BOMD, the nuclei are propagated classically on the electronic BO  potential energy surface. The numerical accuracy is subject to the convergence criteria for the SCF procedure and the MD time step. \cite{marx_ab_2009} After choosing initial positions and velocities for all nuclei, the ground-state energy and forces are computed at the chosen level of theory, and the dynamics of the nuclei are propagated classically according to 
\begin{equation}\label{eq:classical_nuclei}
    M_A \ddot{\mb{R}}_A = -\nabla_A U_{\mr{DFT}},
\end{equation}
where $M_A$ is the mass of the $A$-th nucleus and $\nabla_A U_{\mr{DFT}}$ is the gradient of the ground-state energy $U_{\mr{DFT}}$ with respect to the Cartesian coordinates of the $A$-th nucleus. 

The NEO-BOMD approach is algorithmically similar to conventional 
BOMD, but the classical nuclei move on the electron-proton vibronic surface rather than the electronic surface. The BO separation between the electrons and quantum protons is \textit{not} invoked, but there is a BO separation between the quantum subsystem, composed of the electrons and quantized protons, and the classical nuclei. In other words, the electrons and quantum protons respond instantaneously to the motion of the classical nuclei on the ground state vibronic surface.  Nonadiabatic effects between the classical nuclei and the quantum subsystem can be incorporated using Ehrenfest\cite{li_ab_2005, tully_ehrenfest_2023} or surface hopping\cite{tully_molecular_1990, hammesschiffer_proton_1994} dynamics on NEO vibronic surfaces, but this work focuses on the dynamics on the ground-state vibronic surface.

If the quantum protons were represented by a complete protonic basis set, the NEO energy, $U_{\mr{NEO}}$, at a given classical nuclear configuration, $\mb{R}$, would be invariant with respect to the protonic basis function center positions. In practice, however, we use finite protonic basis sets that are not complete. Thus, $U_{\mr{NEO}}$ does indeed vary with respect to the protonic basis function center positions. Rigorously, $U_{\mr{NEO}}$ is defined as the energy obtained by variationally optimizing all protonic basis function center positions, leading to the following conditions being satisfied for all points on the NEO vibronic surface:
\begin{equation}\label{eq:neo_pes_condition}
    \frac{\partial U_{\mr{NEO}}}{\partial\mc{R}_I} = 0, \hspace{0.5em}\forall I\in\{1,\dots,N^{\mr{p}}\}.
\end{equation}
Thus, in conventional BOMD, only one SCF procedure and force evaluation is required at each time step to obtain the ground-state energy and gradient, respectively, but in NEO-BOMD, the protonic basis function centers must be optimized at each time step, requiring many energy and force evaluations per time step. This necessity greatly inflates the cost of running a NEO-BOMD trajectory. Moreover, occasionally the proton basis function center optimization leads to unstable or local minima, causing  numerical issues with propagation of the equations of motion.

In order to avoid optimizing the protonic basis function centers at each time step, their positions can be represented as additional degrees of freedom and propagated classically according to Eq.~\eqref{eq:classical_nuclei}, with fictitious masses equal to that of a proton. This strategy has been implemented previously in the context of real-time NEO dynamics methods,\cite{zhao_real-time_2020, zhao_nuclear-electronic_2020, xu_lagrangian_2024, li_energy_2025} but not yet in the context of adiabatic dynamics on the NEO ground-state vibronic surface. In this case, the NEO Lagrangian, 
\begin{equation}\label{eq:neo_lagr}
    \mc{L_{\mr{NEO}}} = \frac{1}{2}\sum_A^{N^{\mr{c}}} M_A\lvert\dot{\mb{R}}_A\rvert^2 - U_{\mr{NEO}}(\mb{R}),
\end{equation}
is extended according to 
\begin{equation}\label{eq:ext_neo_lagr}
\mc{L}_{\mr{NEO}}^{\mr{ext}} = \frac{1}{2}\sum_A^{N^{\mr{c}}} M_A\lvert\dot{\mb{R}}_A\rvert^2 + \frac{1}{2}\sum_I^{N^{\mr{p}}} m_{\mr{p}}\lvert\dot{\mc{R}}_I\rvert^2- U_{\mr{NEO}}^{\mr{eff}}(\mb{R},\bm{\mc{R}}).
\end{equation}
In Eq.~\eqref{eq:neo_lagr}, $U_{\mr{NEO}}$ refers to the NEO vibronic surface obtained when Eq.~\eqref{eq:neo_pes_condition} is satisfied for all classical nuclear coordinates $\mb{R}$. In Eq.~\eqref{eq:ext_neo_lagr}, $U_{\mr{NEO}}^{\mr{eff}}$ refers to the NEO energy computed at $\mb{R}$ with the protonic basis function center positions at $\bm{\mc{R}}$, where Eq.~\eqref{eq:neo_pes_condition} is not necessarily satisfied. Thus, $U_{\mr{NEO}}^{\mr{eff}}$ can be viewed as an ``effective'' potential energy of the extended system composed of the classical nuclear coordinates and the protonic basis function center coordinates. We emphasize that in the limit of a complete basis, $U_{\mr{NEO}} = U_{\mr{NEO}}^{\mr{eff}}$ for all $\bm{\mc{R}}$ and $\mb{R}$.

The NEO-ELMD approach presented herein corresponds to the dynamics based on the extended NEO Lagrangian of Eq.~\eqref{eq:ext_neo_lagr}. In this approach, only one NEO-SCF and one force evaluation are performed per time step, as in conventional BOMD. NEO-ELMD is systematically improvable in that it will always approach the NEO-BOMD trajectory obtained by satisfying Eq.~\eqref{eq:neo_pes_condition} as the protonic basis set approaches completeness. NEO-ELMD can also be expected to qualitatively reproduce the NEO-BOMD trajectory assuming that the classical propagation of the protonic basis function centers closely follows the evolution of their variationally optimized positions. As with all extended Lagrangian approaches, NEO-ELMD does not conserve the ``physical'' energy of the system, $E_{\mr{phys}}$, which is defined as the sum of the kinetic energy of the classical nuclei and the NEO potential energy $U_{\mr{NEO}}$. Instead, NEO-ELMD conserves the ``extended'' energy of the system, $E_{\mr{ext}}$, defined as the sum of the kinetic energy of the classical nuclei, the kinetic energy associated with the protonic basis function centers, and the NEO effective potential energy $U_{\mr{NEO}}^{\mr{eff}}$. 

NEO-ELMD propagates the classical nuclei on an extended NEO vibronic surface that is a function of both $\mb{R}$ and $\bm{\mc{R}}$. If CNEO-DFT is performed at each time step instead of NEO-DFT, the extended NEO surface has the same dimensionality, but the basis function center positions $\mc{R}_I$ are replaced with their respective expectation values of the quantum proton position operator, $\expval{\mb{r}_I}$, which are constrained to be equivalent. This extended NEO surface,  where the dependence on $\bm{\mc{R}}$ has been replaced by the dependence on the expectation values of the quantized proton position operator, can be regarded as a constrained minimized energy surface, \cite{chen_incorporating_2023} and dynamics on this surface is the basis of CNEO-MD.\cite{xu_molecular_2022} Thus, a NEO-ELMD trajectory that enforces the constraints in Eq.~\eqref{eq:cneo_constraints} on the NEO-DFT solution at each time step is equivalent to CNEO-MD.

\subsection{Density Matrix Extrapolation and Other Enhancements}\label{subsec:extrap}

BOMD simulations can be made more efficient by speeding up the SCF procedure performed at each time step, often by enhancing the fidelity of the initial guess so that the SCF procedure requires fewer iterations to converge. Various Fock matrix and density matrix extrapolation schemes have been proposed for this purpose. \cite{pulay_fock_2004, polack_grassmann_2021} Although NEO-ELMD is already much cheaper than NEO-BOMD because it requires only one NEO-SCF procedure and force evaluation at each time step, such extrapolation schemes promise to offer additional efficiency gains that can increase the timescale and system size over which NEO-ELMD can be applied.

In this work, we implemented a density matrix extrapolation scheme  for both electronic and protonic densities based on the always stable predictor-corrector (APSC) scheme \cite{kolafa_numerical_1996, kolafa_time-reversible_2004} previously adapted to conventional BOMD. \cite{kuhne_efficient_2007} The concept underlying this extrapolation procedure is to use the history of previously converged density matrices to predict the converged density matrix at the next time step. 

The density matrix extrapolation process that we have implemented is as follows. Let $\mb{P}^{\mr{e(p)}}$ be the electronic (protonic) density matrix in the nonorthogonal atomic orbital basis: $\mb{P}^{\mr{e(p)}} = \mb{C}_{\mr{occ}}^{\mr{e(p)}}\mb{C}_{\mr{occ}}^{\mr{e(p)}^{\mr{T}}}$, where $\mb{C}_{\mr{occ}}^{\mr{e(p)}}$ is the occupied block of the electronic (protonic) coefficient matrix from Eq.~\eqref{eq:roothan}. Let $t_n$ represent the time at the $n$-th time step of a NEO-ELMD trajectory and  $K$ be some positive integer representing how many previous time steps to use in the extrapolation scheme. We can approximate the electronic (protonic) density matrix at time $t_n$ by applying a simple transformation matrix to the converged electronic (protonic) density matrix at time $t_{n-1}$ according to
\begin{equation}\label{eq:extrap_den}
    \mb{P}^{\mr{e(p)}}(t_n) \approx \mb{U}^{\mr{e(p)}}(n,K) \times \mb{P}^{\mr{e(p)}}(t_{n-1})
\end{equation}
where the transformation matrix $\mb{U}^{\mr{e(p)}}$ is dependent on the current time step index $n$ and the chosen $K$. We compute the transformation matrix $\mb{U}^{\mr{e(p)}}$ according to
\begin{equation}\label{eq:extrap_transform}
    \mb{U}^{\mr{e(p)}}(n,K) = \sum_{j=1}^{K} w(j,K) \times \mb{P}^{\mr{e(p)}}(t_{n-j}) \times \mb{S}^{\mr{e(p)}}(t_{n-j}),
\end{equation}
where $\mb{S}^{\mr{e(p)}}$ is the electronic (protonic) atomic orbital overlap matrix and $w(j,K)$ is a coefficient for each term in the summation given by
\begin{equation}\label{eq:extrap_weight}
    w(j,K) = (-1)^{j+1} \times j \times \frac{\binom{2K}{K-j}}{\binom{2K-2}{K-1}}.
\end{equation}
The transformation matrix in Eq.~\eqref{eq:extrap_transform} is computed according to the previous contra-covariant density matrices $\mb{P^{\mr{e(p)}}S^{\mr{e(p)}}}$ because  these evolve more smoothly in time than the coefficient matrices $\mb{C}^{\mr{e(p)}}$. \cite{kuhne_efficient_2007} The weights given by  Eq.~\eqref{eq:extrap_weight} were chosen in order to guarantee sufficient time-reversibility of the original APSC scheme. \cite{kolafa_numerical_1996, kolafa_time-reversible_2004} 

Eqs.\eqref{eq:extrap_den}--\eqref{eq:extrap_weight} provide a clear algorithm for extrapolating the density matrices at a given  time step based on the previous $K$ converged density matrices. Although the predicted density matrix can be an effective initial guess for an SCF procedure, it does not strictly satisfy the idempotency condition
\begin{equation}\label{eq:idem_cond}
    \mb{P}^{\mr{e(p)}} \times \mb{S}^{\mr{e(p)}} \times \mb{P}^{\mr{e(p)}} = \mb{P}^{\mr{e(p)}}, 
\end{equation}
a necessary condition for an SCF solution. However, the predicted density matrices can be expected to nearly satisfy idempotency. Therefore, after extrapolating the electronic and protonic density matrices, we recursively purify \cite{mcweeny_recent_1960} them according to
\begin{equation}\label{eq:mcweeny_purify}
    \mb{P}_{\mr{pred},\perp}^{\mr{e(p)}} \leftarrow 3\mb{P}_{\mr{pred},\perp}^{\mr{e(p)}^2} -2\mb{P}_{\mr{pred},\perp}^{\mr{e(p)}^3},
\end{equation}
where $\mb{P}_{\mr{pred},\perp}^{\mr{e(p)}}$ is the predicted electronic (protonic) density matrix in the orthogonalized atomic orbital basis. 

We iterate the purification routine until the idempotency error satisfies a specified tolerance. The idempotency error is defined as
\begin{equation}\label{eq:idem_error}
    \epsilon_{\mr{idem}} = \lVert \mb{P}_{\mr{pred}}^{\mr{e(p)}} \times \mb{S}^{\mr{e(p)}} \times \mb{P}_{\mr{pred}}^{\mr{e(p)}} - \mb{P}_{\mr{pred}}^{\mr{e(p)}} \rVert_F,
\end{equation}
where $\lVert\cdot\rVert_F$ denotes the Frobenius norm of a given matrix. Once the error is within the specified tolerance, $\mb{P}_{\mr{pred},\perp}^{\mr{e(p)}}$ is back-transformed into the nonorthogonal atomic orbital basis and used as an initial guess for the SCF procedure at that time step. Herein, we introduce the notation NEO-ELMD$(K)$, where $K$ refers to the number of previous time steps used to extrapolate the electronic and protonic density matrices. We use the same notation for CNEO-MD, which enforces the constraints of Eq.~\eqref{eq:cneo_constraints}. 

In addition to this ``extrapolate-then-purify'' approach for obtaining initial guesses to the electronic and protonic density matrices at each time step, we investigated the use of loosely converged NEO-SCF solutions to accelerate NEO-ELMD simulations further. This approximation entails using the extrapolated density matrices as initial guesses but not solving Eq.~\eqref{eq:roothan} to full self-consistency. This investigation was motivated by the ``corrector'' steps following extrapolation (the ``predictor'' step) from the original ASPC scheme\cite{kolafa_numerical_1996, kolafa_time-reversible_2004} and its adaptation to BOMD.\cite{kuhne_efficient_2007} We provide more details and results regarding the use of loosely converged solutions in Section~\ref{subsec:si_more_trajs_malon} of the Supplementary Material. In brief, we found that this approximation was not useful for our simulations. We emphasize that all results presented in the main text solved Eq.~\eqref{eq:roothan} to full self-consistency at each time step.

\section{Computational Details}\label{sec:comp_details}

\begin{figure*}
  \includegraphics[width=5.95in]{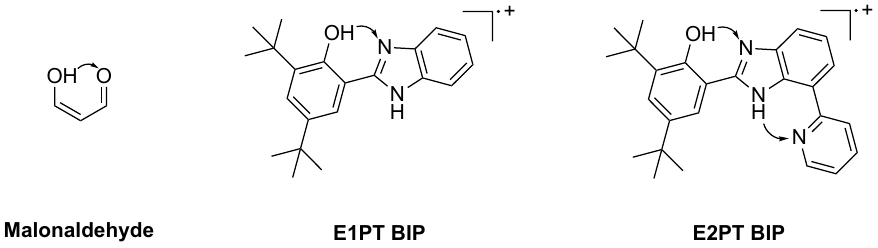} 
  \caption{\justifying Schematic depictions of proton transfers simulated in this work, including the intramolecular proton transfer of malonaldehyde (left), the single proton transfer in E1PT BIP following oxidation (middle), and the double proton transfer relay in E2PT BIP following oxidation (right). Arrows point in the direction of proton transfer. Note that the phenolic proton transfer occurs first in the E2PT system, followed by the imidazolic proton transfer.}
  \label{fig:lewis_structs}
\end{figure*}

In this work, we benchmarked the NEO-ELMD$(K)$ method by simulating the intramolecular proton transfer of malonaldehyde and comparing it to NEO-BOMD as well as to CNEO-MD$(K)$. We then used NEO-ELMD$(K)$ to simulate the nonequilibrium dynamics of the PCET reactions in the  benzimidazole-phenol (BIP) systems exhibiting either one proton transfer (E1PT) or two proton transfers (E2PT), as studied previously with conventional BOMD. \cite{goings_nonequilibrium_2020} The structures and mechanisms of these reactions are depicted in Fig.~\ref{fig:lewis_structs}. The computational details for initializing and propagating these trajectories are given in this section. Further details are given in Section~\ref{sec:si_misc_info}. All methods discussed in this work have been implemented in a developer version of the Q-Chem 6.3 electronic structure software package. \cite{epifanovsky_software_2021}

\subsection{Malonaldehyde Trajectories}\label{subsec:comp_malon}

In all NEO calculations of malonaldehyde, only the transferring proton was treated quantum mechanically, and it was represented by one protonic basis function center. These NEO calculations used the $\omega$B97X\cite{chai_systematic_2008} electron exchange-correlation functional and the epc17-2\cite{brorsen_multicomponent_2017,yang_development_2017} electron-proton correlation functional. These calculations also used the def2-SVP and def2-TZVP\cite{weigend_balanced_2005} electronic basis sets for the classical nuclei and quantum proton, respectively, and a range of protonic basis sets. To obtain the initial coordinates and velocities for the classical nuclei, the equilibrium and transition state structures were optimized at the conventional $\omega$B97X/def2-SVP level of theory.\cite{chai_systematic_2008, weigend_balanced_2005} The protonic basis function center was then optimized at the NEO-DFT level while keeping the classical nuclei at their positions in the equilibrium structure. Starting at this structure, the classical nuclei were assigned initial velocities corresponding to a temperature of 110 K in the direction toward  their positions in the optimized transition state structure. The protonic basis function center was initialized at its optimized position with zero velocity for all trajectories. 

To benchmark the NEO-ELMD$(K)$ method, we varied the protonic basis set (PB4-F2, PB5-G, PB6-H,\cite{yu_development_2020} and even-tempered 8s8p8d8f8g basis\cite{yang_development_2017}), the value of $K$ in the extrapolation routine (0, 2, 4, and 6), the MD time step (0.2 fs, 0.5 fs, and 1.0 fs), and the SCF convergence criterion ($10^{-6}$ $E_{\mr{h}}$ and $10^{-8}$ $E_{\mr{h}}$). Note that $K = 0$ denotes using the previously converged densities as initial guesses following purification. These NEO-ELMD$(K)$ trajectories were compared to corresponding NEO-BOMD trajectories, where the protonic basis function center was optimized variationally at each time step instead of moving classically. In addition, for all NEO-ELMD$(K)$ trajectories propagated, the corresponding CNEO$(K)$ trajectory was also propagated. 

At the start of each malonaldehyde trajectory, an initial single-point calculation was performed utilizing the stepwise NEO-SCF algorithm with the GDM\cite{van_voorhis_geometric_2002} solver. The resulting densities were used as the starting guess for the first time step of the trajectory, and the simultaneous DIIS NEO-SCF\cite{pulay_convergence_1980, pulay_improved_1982, liu_simultaneous_2022} solver was used for the remainder of the trajectory. All malonaldehyde trajectories were propagated for 60 fs.

\subsection{BIP Trajectories}\label{subsec:comp_bip}

The E1PT and E2PT BIP systems\cite{goings_nonequilibrium_2020} are known to exhibit proton transfer upon electrochemical or photochemical oxidation.\cite{odella_controlling_2018,odella_proton-coupled_2019,odella_proton-coupled_2020, yoneda_electronnuclear_2021, arsenault_concerted_2022} To simulate the nonequilibrium dynamics of these systems following oxidation, we optimized structures in the neutral singlet state but propagated them in the cationic doublet state. We propagated NEO-ELMD$(4)$ and conventional BOMD trajectories for both systems, as well as a CNEO-MD$(4)$ trajectory for the E1PT system. 

The E1PT and E2PT neutral singlet systems were optimized at the conventional B3LYP-D3(BJ)/6-31G$^{**}$\cite{hehre_selfconsistent_1972, hariharan_influence_1973, vosko_accurate_1980, lee_development_1988, becke_density-functional_1993, grimme_consistent_2010, grimme_effect_2011} level of theory. For the conventional BOMD trajectories, the nuclei were initialized at the optimized neutral singlet structures and then  were propagated on the cationic doublet surface at the same level of theory using sixth-order Fock matrix extrapolation\cite{pulay_fock_2004} with twelve previous time points. For the NEO-ELMD$(4)$ and CNEO-MD$(4)$ trajectories, the protonic basis function centers for each transferring proton were optimized with NEO-DFT or CNEO-DFT, respectively, in the cationic doublet state with the classical nuclei fixed at the neutral singlet geometry optimized with conventional DFT. All NEO BIP simulations used the B3LYP-D3(BJ)\cite{grimme_consistent_2010, grimme_effect_2011} electron exchange-correlation functional and the epc17-2\cite{brorsen_multicomponent_2017,yang_development_2017} electron-proton correlation functional, as well as the 6-31G$^{**}$ and 6-311G$^{**}$\cite{hehre_selfconsistent_1972, hariharan_influence_1973} electronic basis sets for the classical nuclei and quantum protons, respectively. Note that the D3(BJ) dispersion terms were computed based on the positions of the proton basis function centers. The resulting geometries were used as the initial coordinates for all BIP trajectories.

Starting at these structures for the E1PT and E2PT systems, the classical nuclei were assigned initial velocities according to the following procedure. First, both systems were perturbed along the normal mode corresponding to the bending motion between the phenol and benzimidazole (denoted the ``phenol-benzimidazole bend'') in their neutral singlet states. The displacement vector along this mode was used as the initial velocity vector for each classical nucleus in the system after scaling these velocities to a target temperature of 40 K and 400 K for the E1PT and E2PT systems, respectively. For all NEO trajectories, the protonic basis function centers were initialized at their optimized positions with zero velocities. For the conventional BOMD trajectories, the initial velocities of the transferring protons were set to zero to allow a comparison to the NEO trajectories. This step made no appreciable difference to the conventional BOMD dynamics (see Section~\ref{subsec:si_more_trajs_bip}). In addition to these trajectories, we propagated trajectories for the E1PT system with the initial velocities of all classical nuclei set to zero.

At the start of each trajectory, an initial single-point calculation was performed utilizing the stepwise GDM NEO-SCF algorithm. The resulting electronic and protonic densities were used as the starting guess for the first time step of the trajectory. The E1PT and E2PT NEO trajectories utilized the simultaneous DIIS\cite{pulay_convergence_1980, pulay_improved_1982, liu_simultaneous_2022} and GDM\cite{van_voorhis_geometric_2002, chow_efficient_2026} NEO-SCF algorithms, respectively. A time step of 0.5 fs and convergence criteria of $10^{-7}$ $E_{\mr{h}}$ were used for all BIP trajectories. The  E1PT and E2PT trajectories were propagated for   100 fs and 500 fs, respectively.

We emphasize that the BIP dynamics simulated herein are not comparable to experiment. Such a comparison would require propagating a large number of trajectories with appropriate sampling of initial conditions at finite temperature. The simulations here are rather meant to demonstrate the capabilities of the NEO-ELMD$(K)$ approach in terms of timescale and system size. The comparisons between conventional and NEO trajectories for the initial conditions chosen demonstrate that the qualitative dynamics can change when the transferring protons are quantized. 

\begin{figure*}
  \includegraphics[width=5.95in]{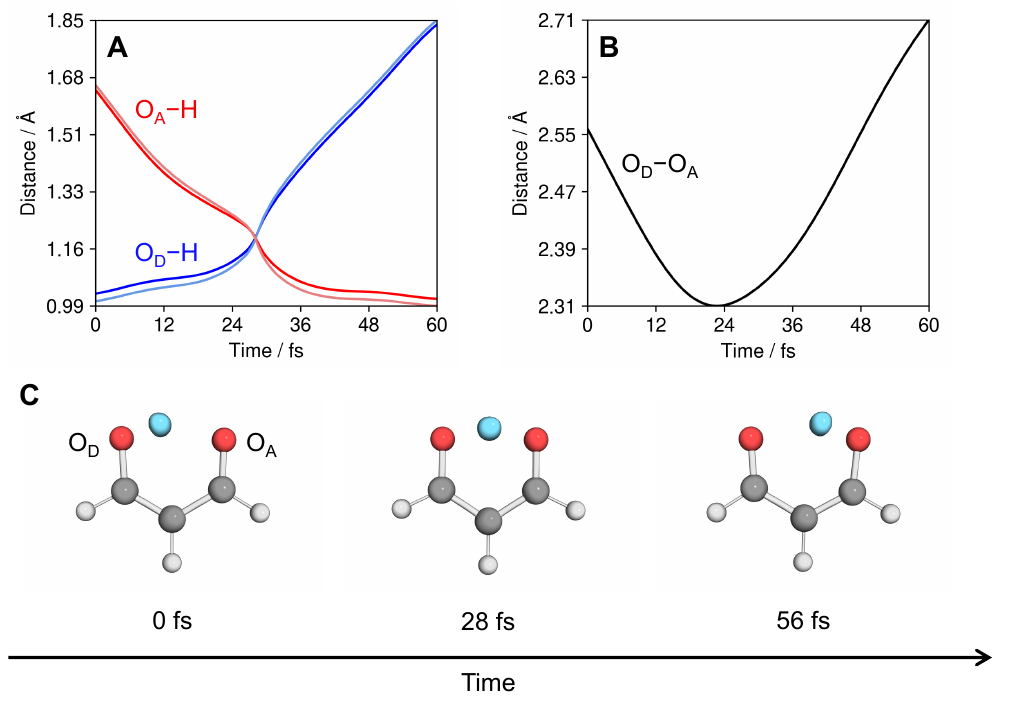} 
  \caption{\justifying NEO-BOMD malonaldehyde trajectory used as a reference for the NEO-ELMD$(K)$ and CNEO-MD$(K)$ trajectories. (A) Distance between the oxygen donor and proton position operator expectation value (blue) and proton basis function center (light blue) and distance between the oxygen acceptor and the proton position operator expectation value (red) and proton basis function center (light red) along the trajectory. (B) Distance between the donor and acceptor oxygens along the trajectory. (C) Evolution of the proton density (shown in cyan with an isosurface value of 0.04 a.u.) along the trajectory.}
  \label{fig:bomd_ref}
\end{figure*}

\subsection{Extrapolation and Optimization Procedures}\label{subsec:comp_extrap}

In this subsection, we provide additional computational details about the density matrix extrapolation and purification procedures described in Section~\ref{subsec:extrap} and  the protonic basis function center optimization mentioned in Section~\ref{subsec:comp_malon}. 

Regarding density matrix extrapolation and purification procedures,\cite{mcweeny_recent_1960} after the electronic and protonic density matrices at each time step were extrapolated according to Eqs.\eqref{eq:extrap_den}--\eqref{eq:extrap_weight}, we recursively purified them according to Eq.~\eqref{eq:mcweeny_purify}. We iterated the purification routine until the idempotency error, $\epsilon_{\mr{idem}}$, defined in Eq.~\eqref{eq:idem_error}, satisfied $\epsilon_{\mr{idem}}\leq10^{-12}$. For time steps $n$ such that $n<K$, where $n=0$ is the initial time step and $K$ is the number of previous time steps used in the extrapolation procedure, we used all previous density matrices for the extrapolation. In other words, we extrapolated the density matrices using all previous time steps for the first $K$ steps. 

Regarding protonic basis function center optimization, we adopted the same convergence criteria used by Q-Chem:\cite{epifanovsky_software_2021} the maximum absolute value of the gradient in any direction on the basis function center must fall below a specified threshold, and either the energy change or displacement of the center from one optimization cycle to the next must fall below a specified threshold. For the gradient tolerance threshold, we tested values of $3.0\times 10^{-4}$ $E_{\mr{h}}a_0^{-1}$, $1.2\times 10^{-4}$ $E_{\mr{h}}a_0^{-1}$, and $3.0\times 10^{-5}$ $E_{\mr{h}}a_0^{-1}$, where $3.0\times 10^{-4}$ $E_{\mr{h}}a_0^{-1}$ is the Q-Chem default value. For both the energy and displacement tolerances, we used the Q-Chem default values of $1.0\times 10^{-8}$ $E_{\mr{h}}$ and $1.2\times 10^{-3}$ $a_0$, respectively.

\section{Results and Discussion}\label{sec:results}

In this section, we present and discuss the NEO-ELMD$(K)$ approach by simulating the intramolecular proton transfer of malonaldehyde, followed by simulations of the E1PT and E2PT BIP systems upon oxidation. For simplicity, we define the proton transfer time for a transferring proton as the first time step at which the expectation value of the position operator associated with that quantum proton becomes closer to its acceptor than to its donor.

\subsection{Malonaldehyde Simulations}\label{subsec:results_malon}

In this subsection, we present an analysis of the trajectories propagated using the PB6-H protonic basis set. We provide additional results for other protonic basis sets in Section~\ref{subsec:si_more_trajs_malon}. The results shown and discussed herein are consistent across the protonic basis sets tested. We find that the NEO-ELMD approach is two to three orders of magnitude faster than the NEO-BOMD approach, where the protonic basis function centers are optimized variationally at each time step, depending on the basis sets and tolerances used in the calculations (Tables~\ref{tab:si_bomd_cpu_times} and~\ref{tab:si_abomd_cpu_times}). This speed-up is obtained without adversely impacting the nuclear--electronic dynamics.

We first provide an analysis of the NEO-BOMD trajectory to which we will compare the NEO-ELMD$(K)$ and CNEO-MD$(K)$ trajectories. This NEO-BOMD trajectory used a time step of 0.5 fs, convergence criteria of $10^{-6}$ $E_{\mr{h}}$, and gradient tolerance of $3.0\times 10^{-5}$ $E_{\mr{h}}a_0^{-1}$. Fig.~\ref{fig:bomd_ref}A shows that the proton transfer proceeds with the expectation value of the quantum proton position operator evolving away from its donor and toward its acceptor in a monotonic fashion. At around 24 fs, where the oxygen donor and acceptor reach their shortest distance along the trajectory (Fig.~\ref{fig:bomd_ref}B), the rate at which the expectation value of the quantum proton position operator moves toward its acceptor noticeably increases, and shortly thereafter it reaches its proton transfer time of 28.5 fs. Subsequently, the distance between the expectation value of the proton position operator and the acceptor begins to plateau, but the distance from the donor steadily increases as the donor and acceptor move apart. Fig.~\ref{fig:bomd_ref}C shows the time-evolution of the proton density in this trajectory. 

The NEO-BOMD reference represents a straightforward proton transfer reaction. We emphasize that this particular set of initial conditions was chosen to ensure that hydrogen tunneling contributions\cite{pak_electron-proton_2004} are minimal. Initializing the classical nuclei with velocities directed toward the conventional transition state structure results in a trajectory where the proton density moves downhill without becoming bilobal. Moreover, the evolution of the expectation value of the proton position operator closely follows the evolution of the optimized position of the protonic basis function center (Fig.~\ref{fig:bomd_ref}A). These observations indicate that the quantum proton does not need to be described by multireference methods, such as NEO multireference configuration interaction (NEO-MRCI) \cite{malbon_nuclearelectronic_2025, stein_computing_2025} or NEO multistate DFT (NEO-MSDFT), \cite{yu_nuclear-electronic_2020,dickinson_generalized_2023} as would be necessary if hydrogen tunneling effects were significant. Analysis of energy conservation and the impact of basis set, convergence criteria, and other parameters is provided in  Section~\ref{subsec:si_more_trajs_malon}. 

\begin{figure}
  \includegraphics[width=2.8in]{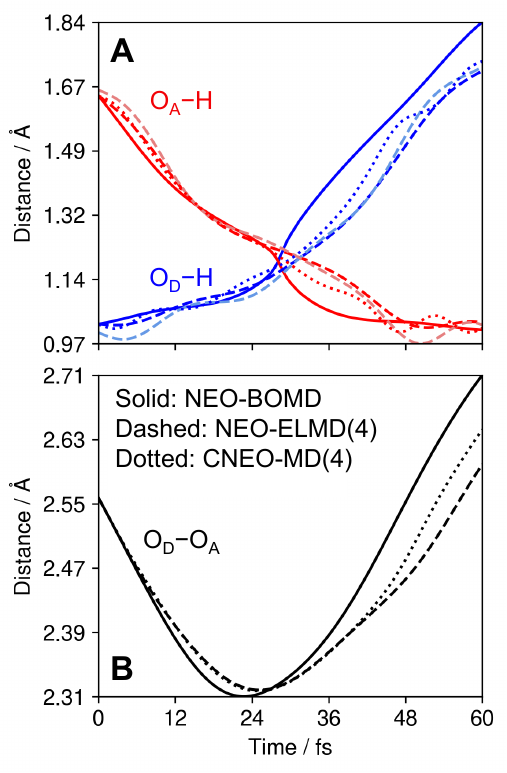} 
  \caption{\justifying NEO-ELMD$(4)$ and CNEO-MD$(4)$ trajectories compared to the NEO-BOMD trajectory for malonaldehyde. NEO-BOMD data is identical to the data shown in Fig.~\ref{fig:bomd_ref}. (A) Distance between the oxygen donor and proton position operator expectation value (blue) and distance between the oxygen acceptor and the proton position operator expectation value (red) along the NEO-BOMD (solid lines), NEO-ELMD$(4)$ (dashed lines), and CNEO-MD$(4)$ (dotted lines) trajectories. The corresponding distances for the proton basis function center are shown in light blue and light red dashed lines for the NEO-ELMD(4) trajectory. (B) Distance between the donor and acceptor oxygens along the NEO-BOMD (solid line), NEO-ELMD$(4)$ (dashed line), and CNEO-MD$(4)$ (dotted line) trajectories. }
  \label{fig:dyn_main}
\end{figure}

Fig.~\ref{fig:dyn_main} shows the NEO-ELMD$(4)$ and CNEO-MD$(4)$ trajectories using the same basis sets, convergence criteria, time step, and other parameters as the NEO-BOMD trajectory shown in Fig.~\ref{fig:bomd_ref}. Recall that the value of $K$ does not impact the dynamics but rather only impacts the efficiency of the simulation. We observe qualitative agreement between both the NEO-ELMD and CNEO-MD trajectories and the NEO-BOMD trajectory, with the NEO-ELMD and CNEO-MD trajectories exhibiting proton transfer times of 32.0 fs and 29.5 fs, respectively. Note that the protonic basis function center position closely follows the proton position operator expectation value in the NEO-ELMD trajectory, as in the NEO-BOMD trajectory. Recall that in CNEO-MD, the proton basis function center and the proton position operator expectation value are constrained to be exactly the same. 

\begin{figure}
  \includegraphics[width=2.8in]{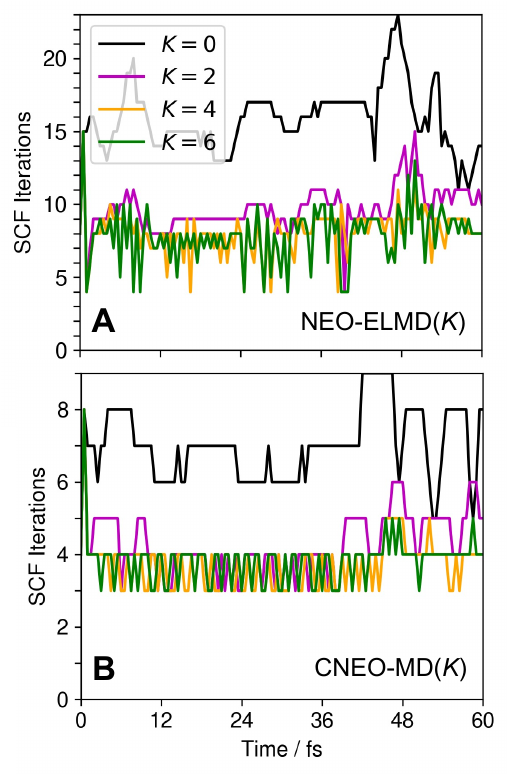} 
  \caption{\justifying Number of iterations of the simultaneous DIIS NEO-SCF procedure needed to fully converge the electronic and protonic densities at each time step at varying values of $K$ for the (A) NEO-ELMD$(K)$ and (B) CNEO-MD$(K)$ malonaldehyde trajectories. $K=0$ indicates that the converged densities from the previous time step were used as initial guesses following purification. All orders of extrapolation tested significantly decrease the number of SCF iterations needed to converge for both the NEO-ELMD and CNEO-MD trajectories, often by a factor of two to four.}
  \label{fig:scf_iter_main}
\end{figure}

\begin{figure}
  \includegraphics[width=2.8in]{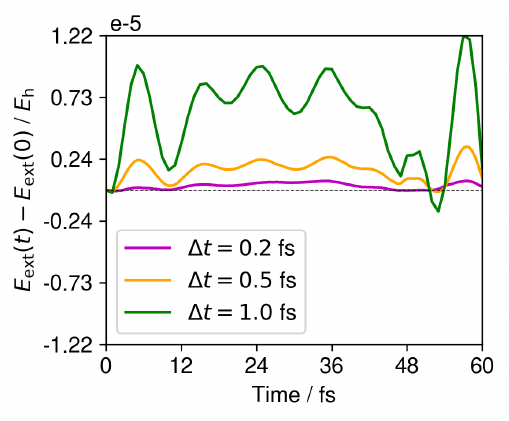} 
  \caption{\justifying Change in extended energy along the NEO-ELMD$(4)$ malonaldehyde trajectories with time steps 0.2 fs (magenta), 0.5 fs (orange), and 1.0 fs (green). As expected, the extended energy is better conserved as the time step decreases.}
  \label{fig:eng_cons_main}
\end{figure}

The CNEO-MD trajectory appears to agree slightly better with the NEO-BOMD trajectory, but this is not a general trend and will depend on the system and level of theory. We note that both the NEO-ELMD and CNEO-MD trajectories exhibit non-monotonic behavior, appearing as O--H vibrations, after proton transfer in Fig.~\ref{fig:dyn_main}A. In the CNEO-MD trajectory, these oscillations would be present even in the limit of a complete protonic basis set, representing the vibrations of the proton position operator expectation value. In NEO-ELMD, however, these vibrations would be less pronounced in the limit of a complete protonic basis because of the underlying assumption of vibronic adiabaticity (i.e., the quantum subsystem responds instantaneously to the classical nuclear motion). 

This observation highlights a fundamental difference between the NEO-ELMD and CNEO-MD approaches. For fully converged basis sets, NEO-ELMD will more faithfully reproduce the results of NEO-BOMD because both NEO-ELMD and NEO-BOMD are propagating the classical nuclei on the ground state adiabatic vibronic surface. In NEO-BOMD, the protonic basis function centers are optimized variationally at each time step, whereas in NEO-ELMD, this optimization is approximated by moving the centers classically on an extended NEO vibronic surface to account for the finite proton basis set. In contrast, CNEO-MD is not performed on the adiabatic vibronic ground state but rather on a constrained minimized energy surface that includes quantum nuclear delocalization effects. In other words, CNEO-MD propagates the dynamics of the classical nuclei and the proton position operator expectation value on the constrained minimized energy surface, which has a different topology than the extended NEO vibronic surface due to the constraints. In this sense, CNEO-MD is not expected to reproduce NEO-BOMD as well as NEO-ELMD does but rather is designed to capture some of the qualitative dynamics of the vibrational motions.

Fig.~\ref{fig:scf_iter_main} shows the number of simultaneous DIIS iterations needed to fully converge the electronic and protonic densities with varying values of $K$ for the NEO-ELMD$(K)$ and CNEO-MD$(K)$ trajectories. Recall that $K=0$ refers to using the converged densities from the previous time step as the initial guess following purification. These results show that the extrapolation procedure significantly decreases the number of SCF iterations needed to converge at each time step for both the NEO-ELMD and CNEO-MD trajectories. For higher values of $K$, the number of iterations needed to converge compared to $K=0$ is often decreased by a factor of two to four. Near the end of the trajectories, the extrapolation procedure offers a more modest speedup. For other protonic basis sets tested, $K=0$ trajectories occasionally resulted in a failure to converge and therefore premature termination, but finite $K$ allowed successful completion of the trajectory (see Section~\ref{subsec:si_more_trajs_malon}). Based on the results of Fig.~\ref{fig:scf_iter_main} and Section~\ref{subsec:si_more_trajs_malon}, we conclude that $K=4$ offers essentially the same level of speedup as $K=6$, and both  these values of $K$ generally outperform the lower $K=2$. Note that CNEO-DFT often requires fewer iterations to converge than NEO-DFT, but this is not always the case, especially for systems with more challenging electronic structures, such as the BIP systems discussed below. 

\begin{figure*}
  \includegraphics[width=6.57in]{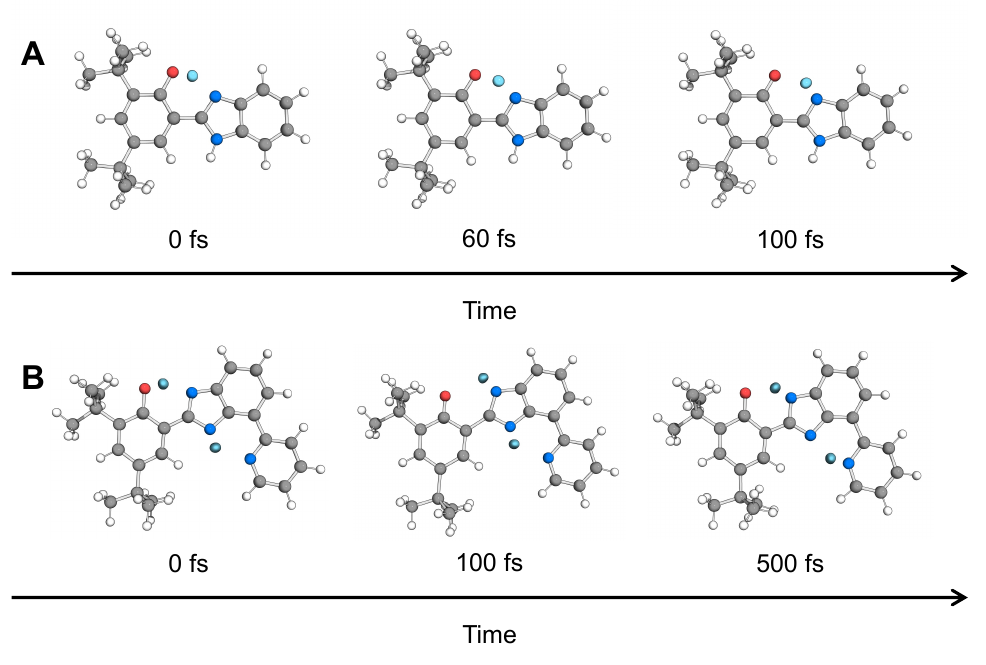} 
  \caption{\justifying Select structures along the NEO-ELMD$(4)$ trajectories for the (A) E1PT and (B) E2PT BIP systems. The initial and final structures of each trajectory are shown, as well as the structures at the proton transfer time for the E1PT system and after the phenolic proton transfer but before the imidazolic proton transfer for the E2PT system. The proton densities are shown in cyan with an isosurface value of 0.04 a.u. The E1PT NEO-ELMD$(4)$ trajectory shown in Panel (A) was initialized with zero velocities.}
  \label{fig:bip_densitites}
\end{figure*}

Finally, we analyze the energy conservation along the NEO-ELMD$(4)$ and CNEO-MD$(4)$ trajectories in Fig.~\ref{fig:eng_cons_main}. As discussed above, the extended energy is defined as the sum of the kinetic energy of the classical nuclei, the kinetic energy associated with the protonic basis function centers for NEO-ELMD or the proton position operator expectation values for CNEO-MD, and the NEO potential energy. A time step of 1.0 fs conserves the extended energy on the order of $10^{-5}$ $E_{\mr{h}}$, and the smaller time steps of 0.5 fs and 0.2 fs conserve the extended energy on the order of $10^{-6}$ $E_{\mr{h}}$ and $10^{-7}$ $E_{\mr{h}}$, respectively.

\subsection{Nonequilibrium Dynamics of BIP Molecules}\label{subsec:results_bip}

We now analyze the results of our simulations of the E1PT and E2PT BIP systems following instantaneous oxidation, which can occur either electrochemically or photochemically. Based on the results discussed above and the timings of malonaldehyde trajectories provided in Section~\ref{sec:si_timings}, we set $K=4$ for our BIP trajectories. This value of $K$ offers a sufficient speedup while not adversely affecting extrapolation stability (see Tables~\ref{tab:si_abomd_cpu_times} and~\ref{tab:si_bip_cpu_times}) or having a major storage requirement. \cite{kolafa_numerical_1996, kolafa_time-reversible_2004, kuhne_efficient_2007} We provide select structures along NEO-ELMD$(4)$ trajectories for both the E1PT and E2PT systems in Fig.~\ref{fig:bip_densitites}.\cite{pymol}  See Fig.~\ref{fig:si_conv_bip_trajs} for select BIP structures along the conventional BOMD trajectories.

\begin{figure}
  \includegraphics[width=2.8in]{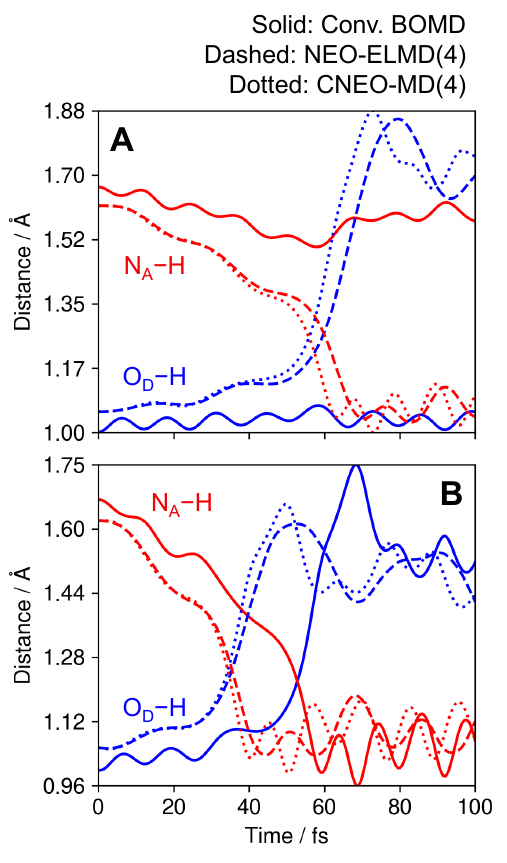} 
  \caption{\justifying Distance between the oxygen donor and the classical proton or proton position operator expectation value (blue) and distance between the nitrogen acceptor and the classical proton or proton position operator expectation value (red) along the conventional BOMD (solid lines), NEO-ELMD$(4)$ (dashed lines), and CNEO-MD$(4)$ (dotted lines) E1PT trajectories, initialized with (A) zero velocities and (B) velocities directed along the phenol-benzimidazole bend mode.}
  \label{fig:e1pt_dyn}
\end{figure}

First we analyze the E1PT system, where we have investigated two sets of initial velocities. The first set corresponds to zero initial velocities for all nuclei, and the second set corresponds to initial velocities directed along the phenol-benzimidazole bend mode. Fig.~\ref{fig:e1pt_dyn} shows the distances between the oxygen donor and nitrogen acceptor and either the classical proton position or the quantum proton position operator expectation value for the conventional BOMD and NEO trajectories. The trajectories obtained with zero initial velocities and non-zero initial velocities are shown in Fig.~\ref{fig:e1pt_dyn}A and B, respectively. Fig.~\ref{fig:si_e1pt_da_dist} shows the distance between the oxygen donor and nitrogen acceptor along each of these trajectories. 

The dynamics differ qualitatively when the proton is treated classically in the conventional BOMD trajectory compared to the quantum treatment in the NEO trajectories when the initial velocities are zero (Fig.~\ref{fig:e1pt_dyn}A). A classical treatment of the transferring proton does not lead to proton transfer under these conditions. In contrast, both the NEO-ELMD$(4)$ and CNEO-MD$(4)$ trajectories exhibit proton transfer with similar dynamics and proton transfer times of 60.0 fs and 56.0 fs, respectively. Interestingly, the oxygen donor and nitrogen acceptor achieve their closest distance to each other at roughly the same time in all three trajectories, at around 50 fs (Fig.~\ref{fig:si_e1pt_da_dist}A). The successful proton transfer in both NEO trajectories is due mainly to the inclusion of the quantum proton zero-point energy in the underlying NEO vibronic surface, which can significantly reduce proton transfer barriers. \cite{schneider_transition_2021} 

The classical and quantum treatments of the transferring proton give more similar results for the trajectories initiated with non-zero velocities (Fig.~\ref{fig:e1pt_dyn}B). In this case, the conventional BOMD, NEO-ELMD$(4)$, and CNEO-MD$(4)$ trajectories exhibit proton transfer with proton transfer times of 53.0 fs, 35.5 fs, and 34.5 fs, respectively. The accelerated proton transfer observed in the NEO trajectories compared to the conventional BOMD trajectory is again due to the inclusion of the quantum proton zero-point energy, which lowers the proton transfer barrier and enhances the rate. 

\begin{figure}
  \includegraphics[width=2.8in]{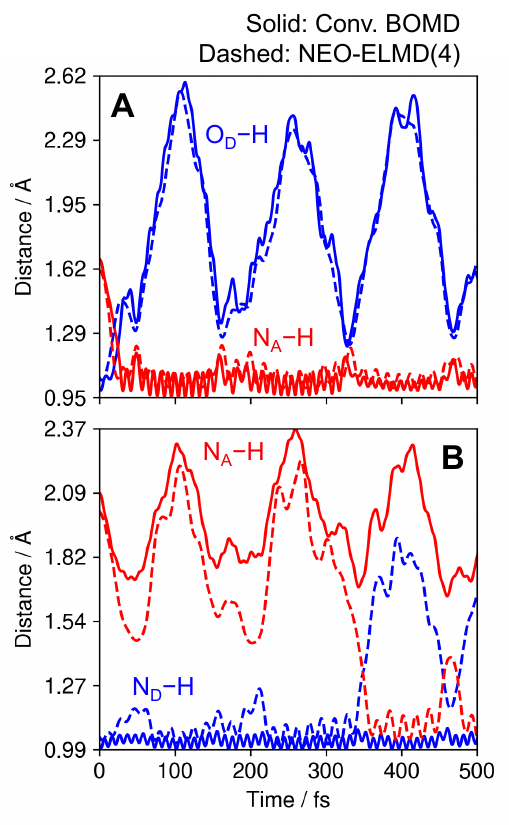} 
  \caption{\justifying Distances between the donor and the classical proton or proton position operator expectation value (blue) and distance between the corresponding acceptor and the classical proton or proton position operator expectation value (red) along the conventional BOMD (solid lines) and NEO-ELMD$(4)$ (dashed lines) E2PT trajectories for the (A) phenolic and (B) imidazolic protons.}
  \label{fig:e2pt_dyn}
\end{figure}

Next, we analyze the E2PT trajectories. For this system, we  propagated  conventional BOMD and NEO-ELMD$(4)$ trajectories with initial velocities along the phenol-benzimidazole bend mode. Oxidation of this molecule leads to proton transfer from the phenol to the imidazole (i.e., from an oxygen to a nitrogen atom), followed by a second proton transfer from the imidazole to the pyridine (i.e., between two nitrogen atoms). Fig.~\ref{fig:e2pt_dyn}A and~\ref{fig:e2pt_dyn}B show the dynamics of the phenolic and imidazolic proton transfer reactions, respectively, each showing the distance between the classical proton positions or the quantum proton position operator expectation values for the conventional BOMD and NEO-ELMD$(4)$ trajectories. Fig.~\ref{fig:si_e2pt_da_dist} shows the distance between the proton donors and acceptors along these trajectories.

Again, we observe a significant difference in the qualitative dynamics obtained with classically treated protons in the conventional BOMD trajectory and with quantum mechanically treated protons in the NEO-ELMD$(4)$ trajectory, particularity for the imidazolic proton transfer. The NEO-ELMD$(4)$ trajectory exhibits asynchronous double proton transfer, with the phenolic and imidazolic proton transfers occurring at 16.5 fs and 349.0 fs, respectively. The conventional BOMD trajectory exhibits the phenolic proton transfer at 24.5 fs but does not show imidazolic proton transfer at all. The dynamics of the phenolic proton transfer are similar in the conventional BOMD and NEO-ELMD$(4)$ trajectories, with the NEO-ELMD$(4)$ trajectory predicting a slightly faster phenolic proton transfer time, as expected due to the inclusion of zero-point energy  effects. Although this difference in phenolic proton transfer times is subtle, the imidazolic proton transfer does not even occur at all without a quantum treatment of the transferring protons. Note that after the imidazolic proton transfer in the NEO trajectory, the proton briefly becomes closer to its donor than its acceptor again at 455.5 fs, but it goes back to being closer to its acceptor a short time later at 475.0 fs. This behavior corresponds to a type of recrossing that occurs prior to stabilization of the final product state.

We emphasize that simulating molecular systems of the size of these BIP molecules over these timescales would have been intractable with NEO-BOMD, particularly NEO-BOMD simulations with tight enough gradient tolerance criteria to yield adequate energy conservation. For a relatively low computational cost, NEO-ELMD allows the simulation of such molecules over these timescales and thereby provides insights about the impact of NQEs on the dynamics of such PCET processes.

\section{Conclusions}\label{sec:conclusions}

In this work, we have introduced an extended Lagrangian approach to NEO molecular dynamics on the adiabatic vibronic ground-state surface. The NEO-ELMD framework employs a similar workflow to conventional BOMD, with one SCF and one force evaluation per time step. The NQEs associated with the quantized protons are incorporated directly into the extended vibronic surface on which the classical nuclear dynamics are performed. The movement of the protonic basis function centers according to classical mechanics allows the use of a relatively localized protonic basis set while maintaining extended energy conservation and offering several orders of magnitude speedup compared to variational optimization at each time step. We introduced density matrix extrapolation as a means of further enhancing the efficiency of these simulations and showed that this scheme results in a notable speedup while maintaining the accuracy of the dynamics. This technique is also applicable to CNEO-MD simulations, resulting in similar computational speedups.

With the methods introduced herein, we are now able to simulate proton transfer dynamics with quantum protons for molecular systems of a significantly larger size and over longer timescales than were previously accessible with multicomponent DFT methods. In addition, this work provides a foundation for future methods that include hydrogen tunneling effects\cite{pak_electron-proton_2004} using the NEO multistate DFT approach.\cite{yu_nuclear-electronic_2020,dickinson_generalized_2023} It also provides the foundation for nonadiabatic dynamics simulations using Ehrenfest\cite{li_ab_2005} or surface hopping\cite{tully_molecular_1990} on the vibronic surfaces,\cite{yu_nonadiabatic_2022, dickinson_nonadiabatic_2024} as well as NEO quantum mechanical/molecular mechanical (QM/MM)\cite{chow_nuclearelectronic_2023, chow_nuclearelectronic_2023-1, chow_nuclear_2024} simulations of condensed-phase systems.


\section*{Supplementary Material}
\noindent Further details on trajectory initialization; discussion of energy conservation; additional malonaldehyde and BIP trajectories; timings of select trajectories; Cartesian coordinates of relevant structures. 

\bigskip

\section*{Acknowledgments}
\noindent This method development was supported by the National Science Foundation Grant No. CHE-2408934. 
E.P. acknowledges funding through a Postdoc Mobility fellowship by the Swiss National Science Foundation (SNSF, Project No. 210754) for implementation of CNEO-DFT and nuclear Hartree product into Q-Chem. The authors thank Dr. Kai Cui and Nicholas Boyer for helpful discussions and comments on the manuscript. The authors thank Dr. Chiara Aieta, Dr. Jonathan Fetherolf, Dr. Scott Garner, Tim Duong, Andreas Ghosh, Rowan Goudy, Jack Morgenstein, Millan Welman, and Jang Mok Yoo for useful discussions. 

\bigskip

\section*{Author Declarations}
\subsection*{Conflict of Interest}
\noindent The authors have no conflicts to disclose.

\bigskip

\section*{Data Availability Statement}
\noindent The data that support the findings of this study are available within this article and its supplementary material. Also, all Q-Chem output files and raw data used to support these findings will be openly available on Github at \href{https://github.com/joseph-dickison20/neo_elmd}{https://github.com/joseph-dickison20/neo\_elmd}. 

\bigskip

\clearpage
\setcounter{section}{0} 
\setcounter{equation}{0}
\setcounter{figure}{0}
\setcounter{table}{0}
\setcounter{scheme}{0}

\onecolumngrid

\clearpage
\begin{titlepage}
    \sffamily
    \centering 
    \vspace{1cm}
    {\Large\bfseries
        Supplementary Material: Extended Lagrangian molecular dynamics on vibronic surfaces in the nuclear--electronic orbital framework\par
    }
    \vspace{1.5cm}
    {\normalsize
    \sffamily Joseph A.\ Dickinson\par
    \vspace{0.2cm}
    \normalfont \textit{Department of Chemistry, Yale University, New Haven, CT 06520, USA}\par
    \normalfont \textit{Department of Chemistry, Princeton University, Princeton, NJ 08544, USA}\par
    \vspace{0.8cm}
    \sffamily Mathew Chow\par
    \vspace{0.2cm}
    \normalfont \textit{Department of Chemistry, Yale University, New Haven, CT 06520, USA}\par
    \normalfont \textit{Department of Chemistry, Princeton University, Princeton, NJ 08544, USA}\par
    \vspace{0.8cm}
    \sffamily Eno Paenurk\par
    \vspace{0.2cm}
    \normalfont \textit{Department of Chemistry, Princeton University, Princeton, NJ 08544, USA}\par
    \vspace{0.8cm}
    \sffamily Sharon Hammes-Schiffer$^{\dagger}$\par
    \vspace{0.2cm}
    \normalfont \textit{Department of Chemistry, Yale University, New Haven, CT 06520, USA}\par
    \normalfont \textit{Department of Chemistry, Princeton University, Princeton, NJ 08544, USA}\par
    }
    \vfill
    $^{\dagger}$\href{mailto:shs566@princeton.edu}{\texttt{shs566@princeton.edu}}\par
\end{titlepage}
\setcounter{page}{0}
\clearpage


\renewcommand{\theequation}{S\arabic{equation}}
\renewcommand{\thefigure}{S\arabic{figure}}
\renewcommand{\thetable}{S\Roman{table}}
\renewcommand{\thescheme}{S\Roman{scheme}}
\renewcommand{\thepage}{S\arabic{page}}
\renewcommand{\thesection}{S\arabic{section}}
\renewcommand{\thesubsection}{\thesection\Alph{subsection}}
\makeatletter
\renewcommand{\p@subsection}{}
\makeatother

\onecolumngrid


\section{Further Details on Trajectory Initialization}\label{sec:si_misc_info}

In this section, we provide additional details regarding the trajectories presented in the main paper. Specifically, we provide further information about obtaining the initial velocities via scaling to a target temperature and the normal modes used to obtain these initial velocities. 

Initializing the malonaldehyde and some of the BIP trajectories relied on scaling velocities to match a target temperature. We review this procedure here. For a system of $N_{\mr{atom}}$ atoms, let $\{\mb{d}_i\}$ be a set of vectors such that $\mb{d}_i$ points from the $i$-th nucleus in a given structure toward its position in some other structure. By the equipartition theorem, the kinetic energy of the system, $K$, is related to the temperature, $T$, via
\begin{equation}
    K = \frac{3}{2}N_{\mr{atom}}k_{\mr{B}}T,
\end{equation}
where $k_{\mr{B}}$ is the Boltzmann constant. Our goal is to obtain velocity vectors for each nucleus $\{\mb{v}_i\}$ such that the total kinetic energy of the nuclei is equal to some target temperature $T_{\mr{target}}$. The kinetic energy of the system is related to $\{\mb{v}_i\}$ according to 
\begin{equation}
    K = \frac{1}{2}\sum_{i=1}^{N_{\mr{atom}}} m_i\lvert\mb{v}_i\rvert^2, 
\end{equation}
where $m_i$ is the mass of the $i$-th nucleus. Using a simple scaling factor $\alpha$ such that $\mb{v}_i = \alpha \mb{d}_i$, 
\begin{align}
    \frac{1}{2}\sum_{i=1}^{N_{\mr{atom}}} m_i\lvert\mb{v}_i\rvert^2 & = \frac{3}{2}N_{\mr{atom}}k_{\mr{B}}T_{\mr{target}} \\
    \alpha^2 \sum_{i=1}^{N_{\mr{atom}}} m_i \lvert\mb{d}_i\rvert^2 &= 3N_{\mr{atom}}k_{\mr{B}}T_{\mr{target}} \\
    \alpha &= \sqrt{\frac{3N_{\mr{atom}}k_{\mr{B}}T_{\mr{target}}}{\sum_{i=1}^{N_{\mr{atom}}} m_i \lvert\mb{d}_i\rvert^2}} \label{eq:si_alpha_scaling}.
\end{align}
Thus, the scaling factor $\alpha$ can be computed according to Eq.~\eqref{eq:si_alpha_scaling} to obtain initial velocity vectors for each nucleus according to $\mb{v}_i = \alpha \mb{d}_i$. 

\begin{figure}
  \includegraphics[width=3.37in]{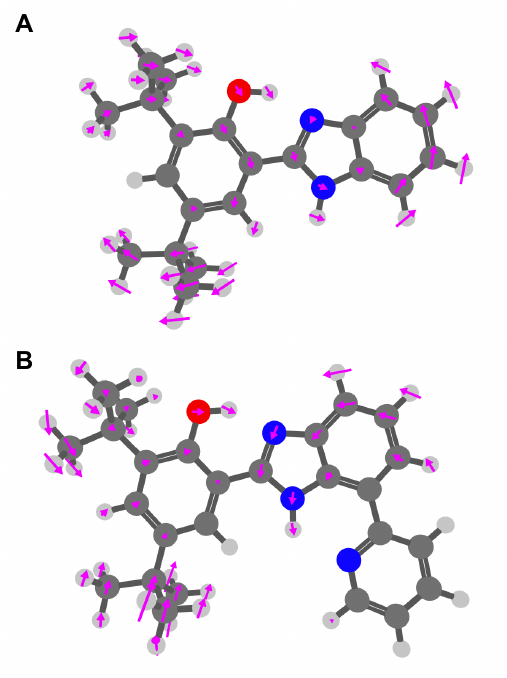} 
  \caption{\justifying (A) E1PT and (B) E2PT structures with normal mode displacement vectors for the phenol-benzimidazole bend mode overlayed on top of their respective nuclei. This bend mode has a frequency of 75.07 cm$^{-1}$ and 240.85 cm$^{-1}$ in the E1PT and E2PT systems, respectively. The magnitudes of the displacement vectors indicate how much a certain nucleus is displaced along the bend mode relative to other nuclei in this normal mode vector.}
  \label{fig:si_bip_norm_modes}
\end{figure}

For initialization of the malonaldehyde trajectories, $\{\mb{d}_i\}$ were unit vectors pointing each nucleus from their position in the equilibrium structure of Table~\ref{tab:si_malon_equil} to their position in  the transition state structure of Table~\ref{tab:si_malon_ts}. For the BIP trajectories with non-zero initial velocities, $\{\mb{d}_i\}$ were the normal mode vectors associated with the phenol-benzimidazole bend mode (see Fig.~\ref{fig:si_bip_norm_modes}). Note that for the malonaldehyde initialization, only the classical nuclei were taken into account in this procedure, but for BIP, all nuclei were included. In the NEO trajectories, the initial velocities of the proton basis function centers were set to zero. After the initial velocities were obtained for the BIP trajectories, the initial velocities of the transferring proton(s) in the conventional BOMD trajectories were set to zero to allow a direct comparison between the conventional BOMD and NEO trajectories. However, we found that retaining non-zero initial velocities for the transferring protons for the conventional BOMD trajectories has only a small impact on the computed proton transfer times (see Figs.~\ref{fig:si_e1pt_pvel_v_no_pvel} and~\ref{fig:si_e2pt_pvel_v_no_pvel}). 

\section{Energy Conservation in NEO-BOMD and NEO-ELMD}\label{sec:si_eng_cons}

In this section, we show that energy conservation in NEO-BOMD depends on terms related to the gradient of the NEO energy with respect to the protonic basis function centers. We also show that the extended energy is rigorously conserved in NEO-ELMD.

Recall that for some fixed configuration of classical nuclei $\mb{R}$, the NEO energy $U_{\mr{NEO}}$ is defined as the minimum energy under which the conditions of Eq.~\eqref{eq:neo_pes_condition} are satisfied. However, as discussed in Section~\ref{subsec:elmd}, $U_{\mr{NEO}}$ is effectively a function of both $\mb{R}$ and the proton basis function center positions $\bm{\mc{R}}$ due to the incompleteness of the protonic basis set. 

Consider $E_{\mr{phys}}$ in NEO-BOMD, defined as the sum of the kinetic energy of the classical nuclei and the NEO potential energy,
\begin{equation}
    E_{\mr{phys}} = \frac{1}{2}\sum_{A=1}^{N^{\mr{c}}} M_A\lvert\dot{\mb{R}}_A\rvert^2 + U_{\mr{NEO}}(\mb{R},\bm{\mc{R}}),
\end{equation}
where we have indicated the dependence of $U_{\mr{NEO}}$ on both $\mb{R}$ and $\bm{\mc{R}}$. Computing the time-derivative of $E_{\mr{phys}}$, we obtain
\begin{equation}
\begin{aligned}
    \dot{E}_{\mr{phys}} &= \sum_{A=1}^{N^{\mr{c}}} M_A \dot{\mb{R}}_A\cdot\ddot{\mb{R}}_A + \sum_{A=1}^{N^{\mr{c}}} \frac{\partial U_{\mr{NEO}}}{\partial \mb{R}_A}\dot{\mb{R}}_A + \sum_{I=1}^{N^{\mr{p}}} \frac{\partial U_{\mr{NEO}}}{\partial \mc{R}_I}\dot{\mc{R}}_I \\
    &= \sum_{A=1}^{N^{\mr{c}}}\left((M_A\ddot{\mb{R}}_A)\cdot\dot{\mb{R}}_A + \frac{\partial U_{\mr{NEO}}}{\partial \mb{R}_A}\dot{\mb{R}}_A\right) + \sum_{I=1}^{N^{\mr{p}}} \frac{\partial U_{\mr{NEO}}}{\partial \mc{R}_I}\dot{\mc{R}}_I \\
    &=  \sum_{A=1}^{N^{\mr{c}}}\left(-\frac{\partial U_{\mr{NEO}}}{\partial \mb{R}_A}\dot{\mb{R}}_A + \frac{\partial U_{\mr{NEO}}}{\partial \mb{R}_A}\dot{\mb{R}}_A\right) + \sum_{I=1}^{N^{\mr{p}}} \frac{\partial U_{\mr{NEO}}}{\partial \mc{R}_I}\dot{\mc{R}}_I \\
    &= \sum_{I=1}^{N^{\mr{p}}} \frac{\partial U_{\mr{NEO}}}{\partial \mc{R}_I}\dot{\mc{R}}_I.
\end{aligned}
\end{equation}
$E _{\mr{phys}}$ is rigorously conserved (i.e., $\dot{E}_{\mr{phys}} = 0$) in NEO-BOMD if any of the following three conditions is satisfied:
\begin{enumerate}
    \item The basis set is complete, and therefore $U_{\mr{NEO}}$ is invariant with respect to the protonic basis function center positions. \label{cond:complete}
    \item All protonic basis function centers are fixed to their initial positions. \label{cond:fixed} 
    \item All protonic basis function center positions are optimized variationally  at all time steps. \label{cond:opt}
\end{enumerate}
Condition~\ref{cond:complete} is not possible to achieve in practice, and Condition~\ref{cond:fixed} is not generally advisable because a computationally tractable fixed protonic basis set is likely too limited in its span of coordinate space to adequately describe proton transfer without knowing the reaction pathway ahead of time. The most general approach is to satisfy Condition~\ref{cond:opt} and optimize all protonic basis function centers at all time steps, but this procedure is computationally expensive, motivating the development of NEO-ELMD in this work. 

In NEO-ELMD, $E_{\mr{ext}}$ is defined as the sum of the kinetic energy of the classical nuclei, the kinetic energy of the protonic basis function centers, and the NEO potential energy, 
\begin{equation}
    E_{\mr{ext}} = \frac{1}{2}\sum_{A=1}^{N^{\mr{c}}} M_A\lvert\dot{\mb{R}}_A\rvert^2 + \frac{1}{2}\sum_{I=1}^{N^{\mr{p}}} M_{\mr{p}}\lvert\dot{\mc{R}}_I\rvert^2 + U_{\mr{NEO}}^{\mr{eff}}(\mb{R},\bm{\mc{R}}),
\end{equation}
where we have indicated that the nuclei evolve on an effective NEO vibronic surface, which is dependent on $\mb{R}$ and $\bm{\mc{R}}$. Computing the time-derivative of $E_{\mr{ext}}$, we obtain
\begin{equation}
\begin{aligned}
    \dot{E}_{\mr{ext}} &= \sum_{A=1}^{N^{\mr{c}}} M_A \dot{\mb{R}}_A\cdot\ddot{\mb{R}}_A + \sum_{A=1}^{N^{\mr{c}}} \frac{\partial U^{\mr{eff}}_{\mr{NEO}}}{\partial \mb{R}_A}\dot{\mb{R}}_A + \sum_{I=1}^{N^{\mr{p}}} M_{\mr{p}} \dot{\mc{R}}_I\cdot\ddot{\mc{R}}_I + \sum_{I=1}^{N^{\mr{p}}} \frac{\partial U^{\mr{eff}}_{\mr{NEO}}}{\partial \mc{R}_I}\dot{\mc{R}}_I \\
    &= \sum_{A=1}^{N^{\mr{c}}}\left((M_A\ddot{\mb{R}}_A)\cdot\dot{\mb{R}}_A + \frac{\partial U^{\mr{eff}}_{\mr{NEO}}}{\partial \mb{R}_A}\dot{\mb{R}}_A\right) + \sum_{I=1}^{N^{\mr{p}}}\left((M_{\mr{p}}\ddot{\mc{R}}_I)\cdot\dot{\mc{R}}_I + \frac{\partial U^{\mr{eff}}_{\mr{NEO}}}{\partial \mc{R}_I}\dot{\mc{R}}_I\right) \\
    &= \sum_{A=1}^{N^{\mr{c}}}\left(-\frac{\partial U^{\mr{eff}}_{\mr{NEO}}}{\partial \mb{R}_A}\dot{\mb{R}}_A + \frac{\partial U^{\mr{eff}}_{\mr{NEO}}}{\partial \mb{R}_A}\dot{\mb{R}}_A\right) + \sum_{I=1}^{N^{\mr{p}}}\left(-\frac{\partial U^{\mr{eff}}_{\mr{NEO}}}{\partial \mc{R}_I}\dot{\mc{R}}_I + \frac{\partial U^{\mr{eff}}_{\mr{NEO}}}{\partial \mc{R}_I}\dot{\mc{R}}_I\right) \\
    &= 0
\end{aligned}
\end{equation}
Thus, $E_{\mr{ext}}$ is rigorously conserved in NEO-ELMD. It is important to note that the physical energy, $E_{\mr{phys}}$, is not conserved in NEO-ELMD. The underlying assumption, as is the case with all extended Lagrangian approaches,\cite{car_unified_1985, marx_ab_2009} is that the exchange of energy between the real and fictitious degrees of freedom in the system is not significant enough to alter the qualitative dynamics. 

\begin{figure*}
  \includegraphics[width=6.69in]{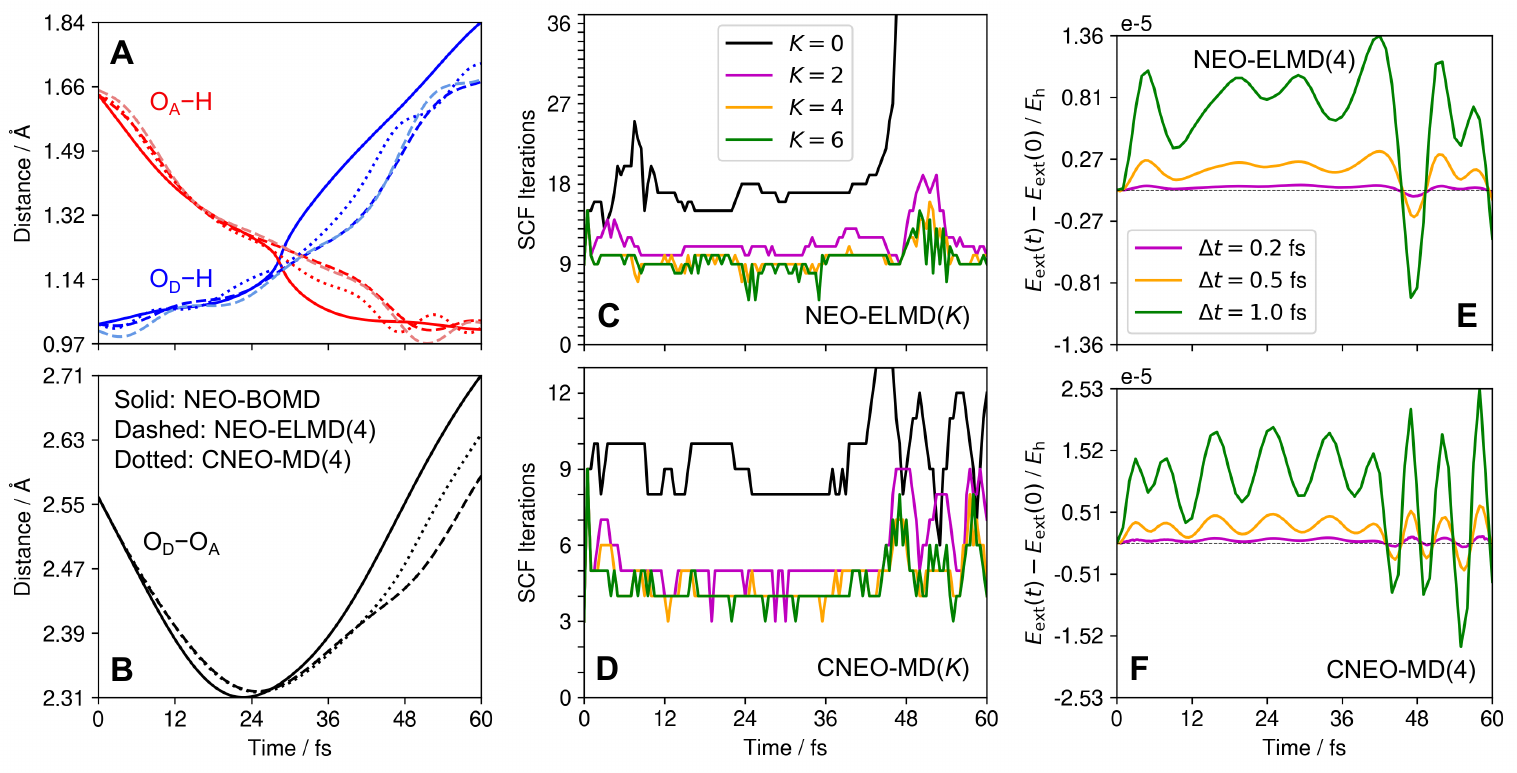} 
  \caption{\justifying Malonaldehyde results with the PB4-F2 protonic basis set. (A) Distance between the oxygen donor and proton position operator expectation value (blue) and distance between the oxygen acceptor and the proton position operator expectation value (red) along the NEO-BOMD (solid lines), NEO-ELMD$(4)$ (dashed lines), and CNEO-MD$(4)$ (dotted lines) trajectories. The corresponding distances for the proton basis function center are shown in light blue and light red dashed lines for the NEO-ELMD(4) trajectory. (B) Distance between the donor and acceptor oxygens along the NEO-BOMD (solid line), NEO-ELMD$(4)$ (dashed line), and CNEO-MD$(4)$ (dotted line) trajectories. (C, D) Number of iterations of the simultaneous DIIS NEO-SCF procedure needed to fully converge the electronic and protonic densities at each time step at varying values of $K$ in the (C) NEO-ELMD$(K)$ and (D) CNEO-MD$(K)$ trajectories. Note that the $K=0$ simulation prematurely terminated due to failed SCF convergence at around 48 fs. (E, F) Change in extended energy along the (E) NEO-ELMD$(4)$ and (F) CNEO-MD$(4)$ trajectories  with time steps 0.2 fs (magenta), 0.5 fs (orange), and 1.0 fs (green).}
  \label{fig:si_dyn_pb4f2}
\end{figure*}

\section{Additional Trajectories and Analysis}\label{sec:si_more_trajs}

In this section, we provide the results of additional trajectories for malonaldehyde and BIP systems. 

\subsection{Malonaldehyde}\label{subsec:si_more_trajs_malon}

\begin{figure*}
  \includegraphics[width=6.69in]{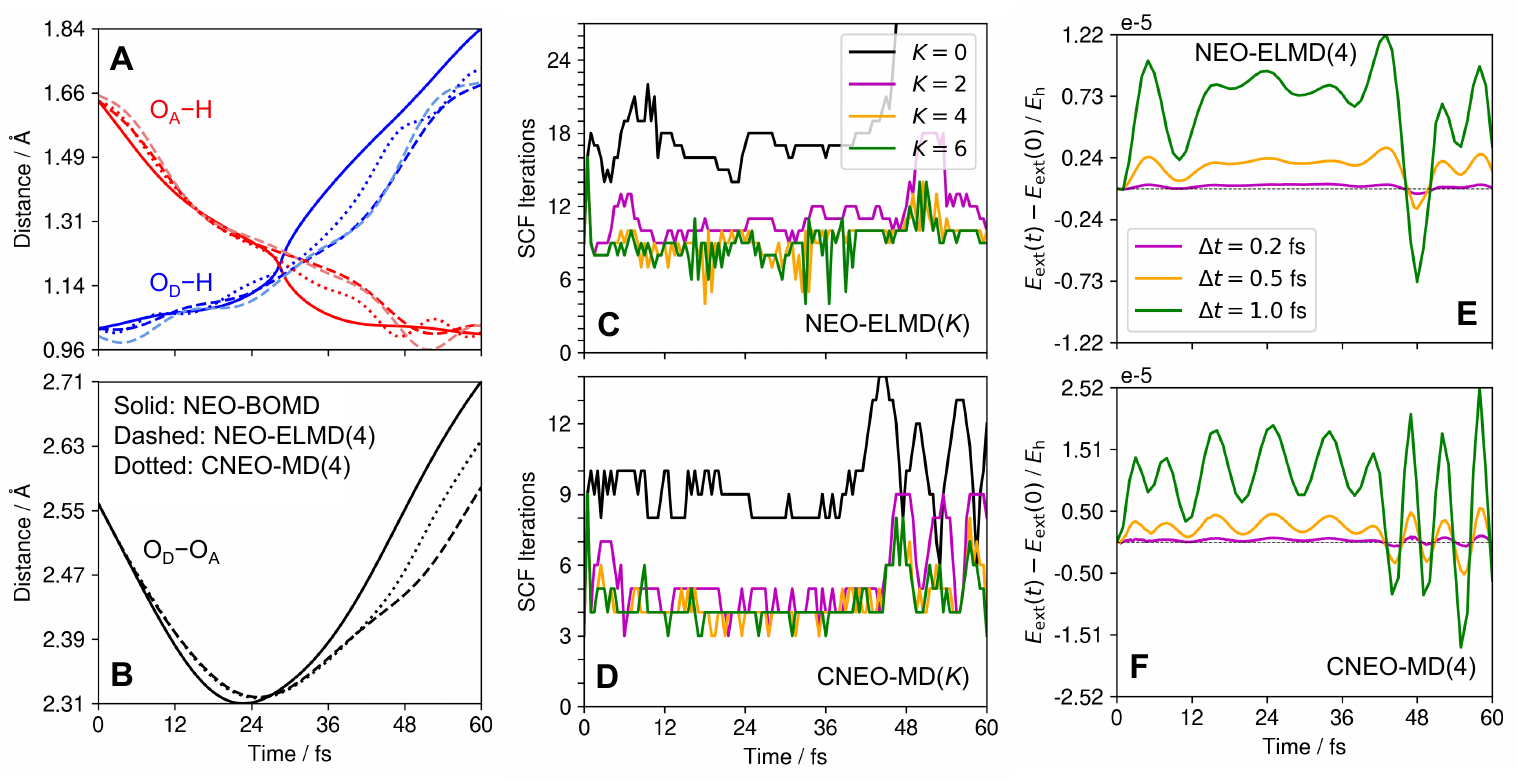} 
  \caption{\justifying Malonaldehyde results with the PB5-G protonic basis set. (A) Distance between the oxygen donor and proton position operator expectation value (blue) and distance between the oxygen acceptor and the proton position operator expectation value (red) along the NEO-BOMD (solid lines), NEO-ELMD$(4)$ (dashed lines), and CNEO-MD$(4)$ (dotted lines) trajectories. The corresponding distances for the proton basis function center are shown in light blue and light red dashed lines for the NEO-ELMD(4) trajectory. (B) Distance between the donor and acceptor oxygens along the NEO-BOMD (solid line), NEO-ELMD$(4)$ (dashed line), and CNEO-MD$(4)$ (dotted line) trajectories. (C, D) Number of iterations of the simultaneous DIIS NEO-SCF procedure needed to fully converge the electronic and protonic densities at each time step at varying values of $K$ in the (C) NEO-ELMD$(K)$ and (D) CNEO-MD$(K)$ trajectories. Note that the $K=0$ simulation prematurely terminated due to failed SCF convergence at around 48 fs. (E, F) Change in extended energy along the (E) NEO-ELMD$(4)$ and (F) CNEO-MD$(4)$ trajectories  with time steps 0.2 fs (magenta), 0.5 fs (orange), and 1.0 fs (green).}
  \label{fig:si_dyn_pb5g}
\end{figure*}

\begin{figure*}
  \includegraphics[width=6.69in]{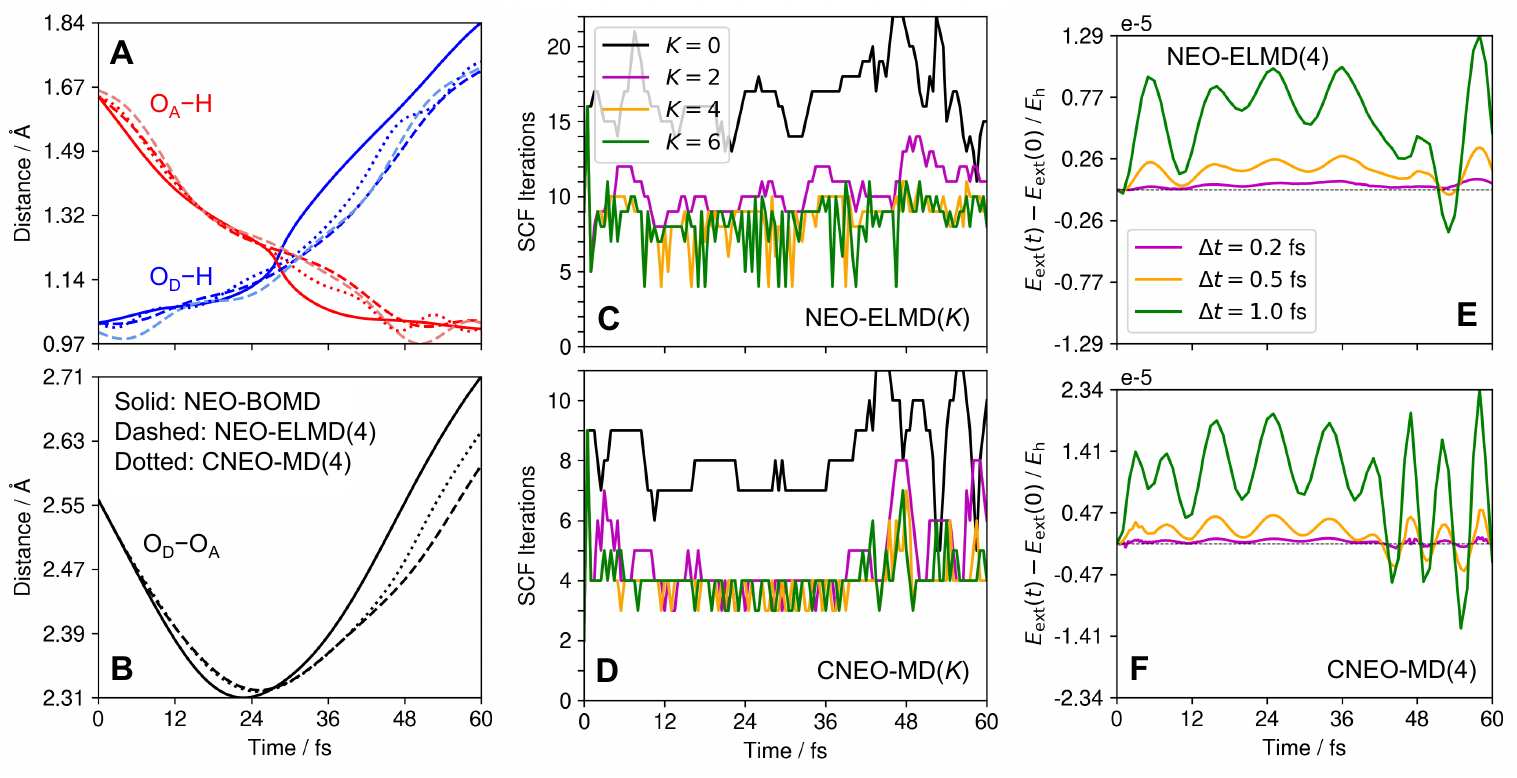} 
  \caption{\justifying Malonaldehyde results with the even-tempered 8s8p8d8f8g protonic basis set. (A) Distance between the oxygen donor and proton position operator expectation value (blue) and distance between the oxygen acceptor and the proton position operator expectation value (red) along the NEO-BOMD (solid lines), NEO-ELMD$(4)$ (dashed lines), and CNEO-MD$(4)$ (dotted lines) trajectories. The corresponding distances for the proton basis function center are shown in light blue and light red dashed lines for the NEO-ELMD(4) trajectory. (B) Distance between the donor and acceptor oxygens along the NEO-BOMD (solid line), NEO-ELMD$(4)$ (dashed line), and CNEO-MD$(4)$ (dotted line) trajectories. (C, D) Number of iterations of the simultaneous DIIS NEO-SCF procedure needed to fully converge the electronic and protonic densities at each time step at varying values of $K$ in the (C) NEO-ELMD$(K)$ and (D) CNEO-MD$(K)$ trajectories. Note that the $K=0$ simulation prematurely terminated due to failed SCF convergence at around 48 fs. (E, F) Change in extended energy along the (E) NEO-ELMD$(4)$ and (F) CNEO-MD$(4)$ trajectories  with time steps 0.2 fs (magenta), 0.5 fs (orange), and 1.0 fs (green).}
  \label{fig:si_dyn_8spdfg}
\end{figure*}

In Figs.~\ref{fig:si_dyn_pb4f2}--\ref{fig:si_dyn_8spdfg}, we show the results of the malonaldehyde trajectories across different protonic basis sets. Note that unless otherwise specified, these simulations use the same parameters as the results given in Section~\ref{subsec:results_malon} of the main text (i.e., 0.5 fs time step and $10^{-6}$ $E_{\mr{h}}$ convergence criteria). Across these basis sets, we see similar results to the PB6-H results given in the main paper: qualitative agreement in the dynamics among NEO-BOMD, NEO-ELMD$(4)$, and CNEO-MD$(4)$, substantial decreases in the number of SCF iterations needed to converge per time step with increasing order of density matrix extrapolation, and sufficient extended energy conservation. Note that all simulations with the even-tempered 8s8p8d8f8g protonic basis used a linear dependence threshold of $10^{-2}$.

\begin{figure*}
  \includegraphics[width=6.69in]{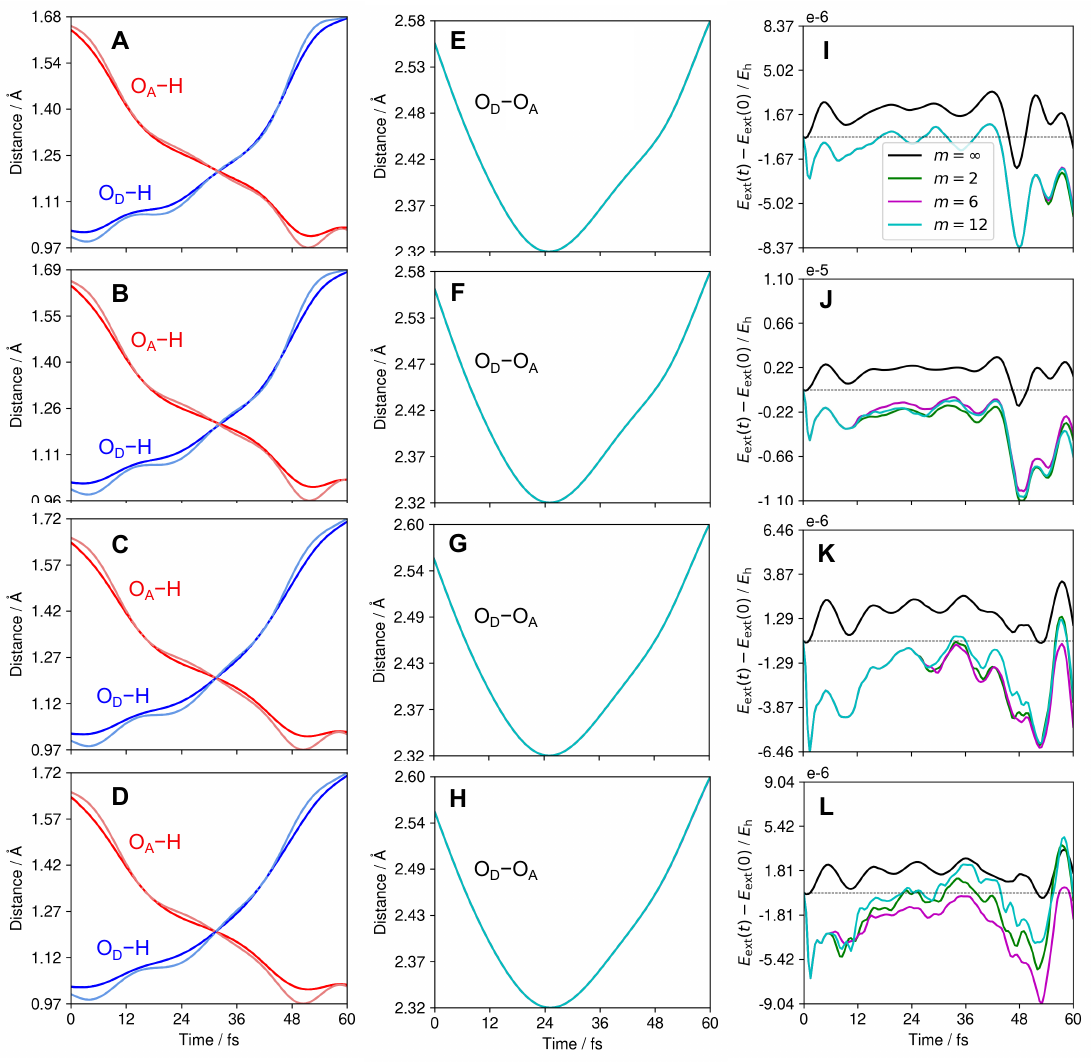} 
  \caption{\justifying Malonaldehyde NEO-ELMD$(4)$ results with different values of $m$ across protonic basis sets, where $m$ is the number of stepwise NEO-SCF macrocycles performed regardless of convergence, and each row corresponds to a different protonic basis set. Panels (A, E , I), (B, F, J), (C, G, K), and (D, H, L) correspond to the protonic basis sets PB4-F2, PB5-G, PB6-H, and even-tempered 8s8p8d8f8g, respectively. Note that $m=\infty$ in this context refers to full convergence, which serves as the reference in these tests. (A--D) Distance between the oxygen donor and proton position operator expectation value (blue) and proton basis function center (light blue) and distance between the oxygen acceptor and the proton position operator expectation value (red) and proton basis function center (light red) along the trajectory. Four different values of $m$, where $m = \{2, 6, 12, \infty\}$, are plotted in each panel from A -- H, but they are not visually distinguishable. (E--H) Distance between the donor and acceptor oxygens along the trajectory. (I--L)  Change in extended energy along the trajectory for varying values of $m$. A fixed value of $m$ results in slightly poorer, but still sufficient, extended energy conservation.}
  \label{fig:si_mvals}
\end{figure*}

Fig.~\ref{fig:si_mvals} shows different malonaldehyde NEO-ELMD$(4)$ results with varying levels of SCF convergence following density matrix extrapolation, again with 0.5 fs time step and $10^{-6}$ $E_{\mr{h}}$ convergence criteria. In these simulations, we performed a specified number of macrocycles of the stepwise NEO-SCF procedure at each time step and allowed the trajectory to proceed regardless of whether the SCF fully converged in that number of macrocycles. See Ref.~\citenum{liu_simultaneous_2022} for a more detailed explanation of stepwise versus simultaneous NEO-SCF procedures. This investigation was motivated by the ``corrector'' steps following extrapolation (the ``predictor'' step) from the original ASPC scheme\cite{kolafa_numerical_1996, kolafa_time-reversible_2004} and its adaption to BOMD.\cite{kuhne_efficient_2007} Let $m$ represent the number of macrocycles performed at each time step of these trajectories. As the plots show, propagating the trajectories with a fixed value of $m$ has a negligible effect on the dynamics. This observation indicates that the extrapolated densities serve as sufficiently good initial guesses to the SCF procedure such that small values of $m$ still yield nearly identical electronic and protonic densities as fully converged solutions. The main impact of a fixed value of $m$ is observed in the extended energy conservation, where lack of full convergence at each time step resulted in slightly greater energy fluctuations, but still on the order of magnitude typically considered acceptable in conventional BOMD. The same trends are observed in the corresponding CNEO-MD$(4)$ trajectories, but these results have been omitted here for brevity. For the systems studied with the computational parameters used herein, the loose convergence did not lead to substantial computational savings, but we anticipate that such an approach would lead to more substantial savings as the expense of each SCF procedure increases. 

\begin{figure*}
  \includegraphics[width=5.0in]{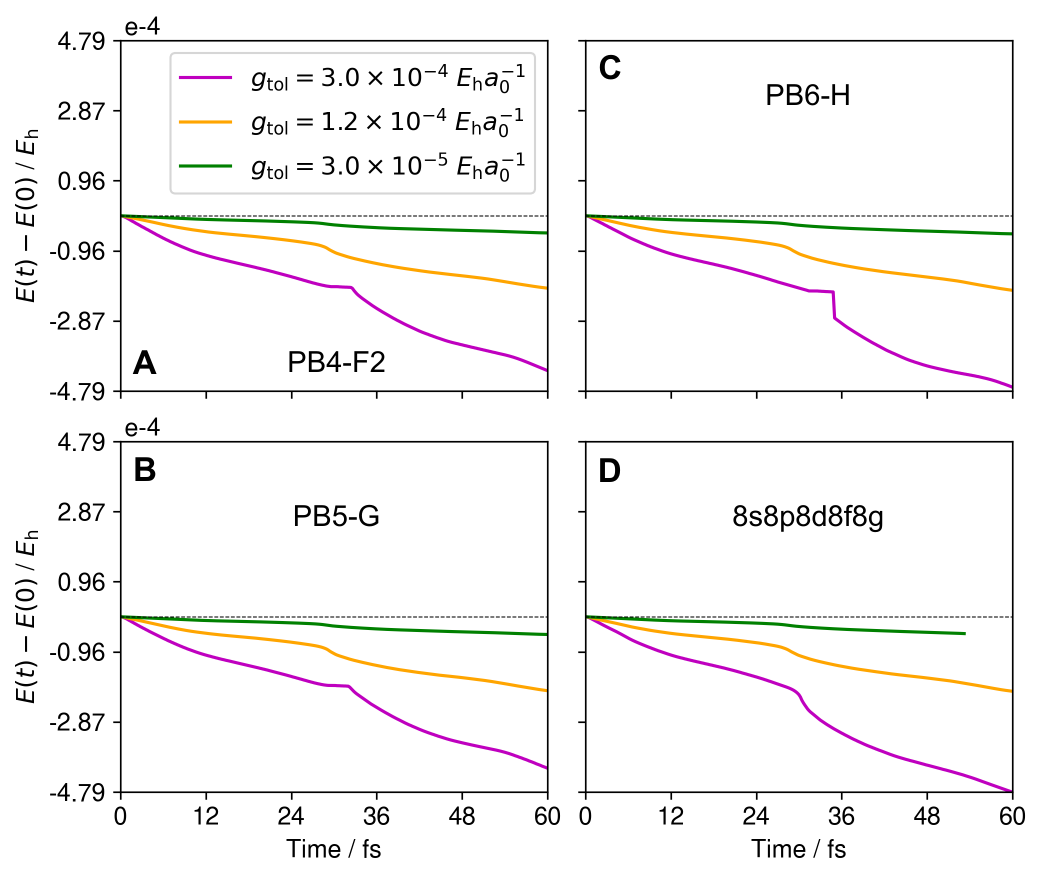} 
  \caption{\justifying Change in total energy along the NEO-BOMD malonaldehyde trajectories with a small time step (0.2 fs) and tight convergence criteria ($10^{-8}$ $E_{\mr{h}}$) at different gradient tolerance thresholds, $g_{\mr{tol}}$, for the protonic basis function center optimization with the protonic basis sets (A) PB4-F2, (B) PB5-G, (C) PB6-H, and (D) even-tempered 8s8p8d8f8g. Note that the 8s8p8d8f8g trajectory with the tightest gradient tolerance was the only trajectory that was not propagated for a full 60 fs due to the computational expense. A gradient tolerance an order of magnitude tighter than the Q-Chem default value is needed in order to conserve energy on the order of $10^{-6}$ $E_{\mr{h}}$.}
  \label{fig:si_bomd_eng_cons}
\end{figure*}

Fig.~\ref{fig:si_bomd_eng_cons} shows the  change in total energy for the NEO-BOMD trajectories with different protonic basis sets as a function of the gradient tolerance for the basis function center optimization at each time step. These trajectories used the smallest time step and strictest convergence criteria tested (0.2 fs and $10^{-8}$ $E_{\mr{h}}$, respectively) in order to isolate the effect of the gradient tolerance on the energy conservation. The results are consistent across basis sets: the Q-Chem\cite{epifanovsky_software_2021} default value for the gradient tolerance is insufficient to conserve energy at what is typically considered to be acceptable in conventional BOMD simulations (i.e., on the order of $10^{-5}$ $E_{\mr{h}}$), and a tolerance an order of magnitude tighter than this value is needed to obtain energy conservation on the order of $10^{-5}$ $E_{\mr{h}}$ or better. This tighter value of the tolerance greatly inflates the number of optimization cycles needed at each time step, making NEO-BOMD with sufficient energy conservation intractable for larger molecular systems. We can understand the monotonic decrease in total energy over time for all of these trajectories as energy lost to the fictitious degrees of freedom associated with the basis function center as it ``moves'' to its optimum position. It is clear that the closer the center is to its variational minimum, the smaller the severity of this monotonic drift. The only way to completely eliminate this drift is to continuously tighten the gradient tolerance, which is computationally intractable. 

\begin{figure}
  \includegraphics[width=2.8in]{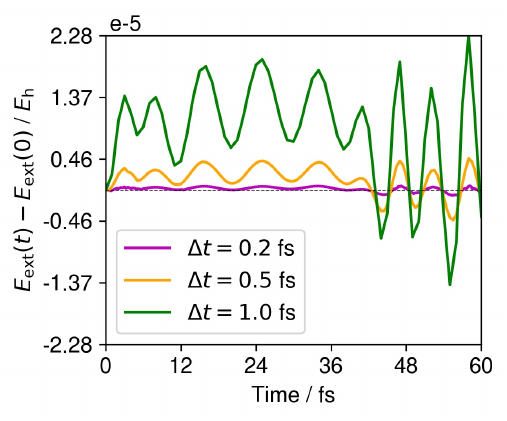} 
  \caption{\justifying Change in extended energy along the CNEO-MD$(4)$ malonaldehyde trajectories with time steps 0.2 fs (magenta), 0.5 fs (orange), and 1.0 fs (green). Note that this figure is analogous to Fig.~\ref{fig:eng_cons_main} for the NEO-ELMD(4) malonaldehyde trajectories in the main paper.}
  \label{fig:si_cneo_eng_cons_main}
\end{figure}

Fig.~\ref{fig:si_cneo_eng_cons_main} shows the energy conservation analysis along  the CNEO-MD$(4)$ trajectory that is analogous to the NEO-ELMD$(4)$ trajectory analyzed in Fig.~\ref{fig:eng_cons_main} of the main paper. As expected, the extended energy is better conserved as the time step decreases. 

\clearpage
\subsection{BIP}\label{subsec:si_more_trajs_bip}

\begin{figure*}
  \includegraphics[width=6.57in]{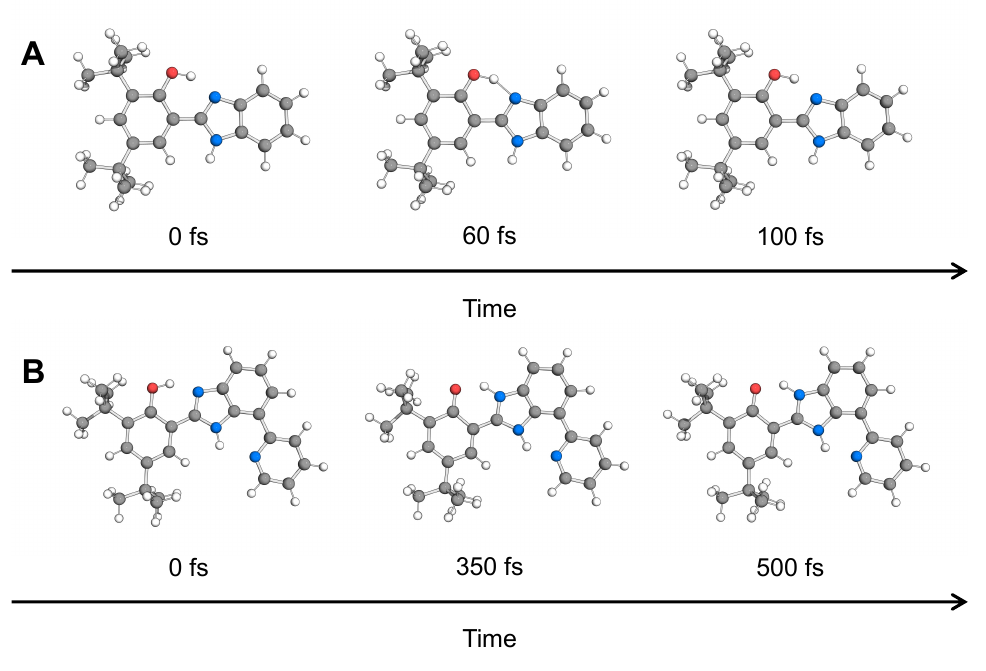} 
  \caption{\justifying Select structures along the conventional BOMD trajectories for the (A) E1PT and (B) E2PT BIP systems. The initial and final structures of each trajectory are shown, as well as the structures at or slightly after the NEO-ELMD$(4)$ proton transfer time for the E1PT system and the NEO-ELMD$(4)$ imidazolic proton transfer time for the E2PT system. The E1PT trajectory shown in Panel (A) was initialized with zero velocities. Note that the transferring classical proton in the middle structure of Panel (A) is close enough to both the oxygen donor and nitrogen acceptor for the visualization software\cite{pymol} to render ``bonds'' to both atoms, but this proton never completes the transfer to the acceptor in this particular trajectory.}
  \label{fig:si_conv_bip_trajs}
\end{figure*}

\begin{figure}
  \includegraphics[width=2.8in]{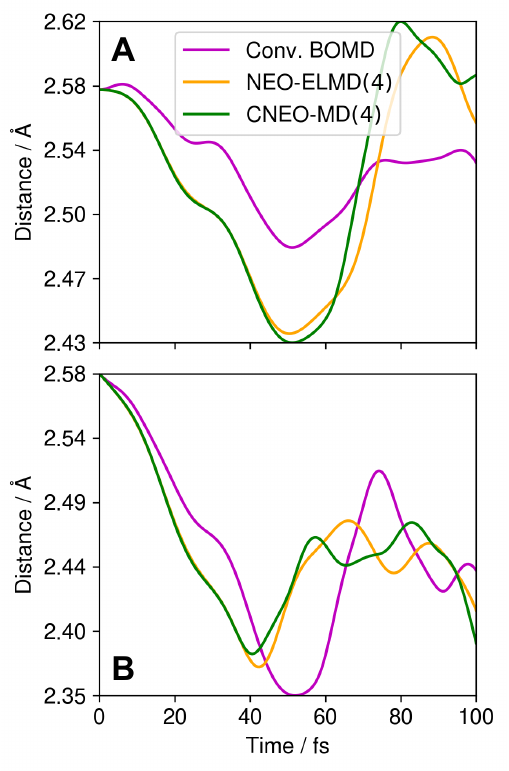} 
  \caption{\justifying Distances between the oxygen donor and nitrogen acceptor along the conventional BOMD (magenta), NEO-ELMD$(4)$ (orange), and CNEO-MD$(4)$ (green) E1PT trajectories, initialized with (A) zero velocities and (B) velocities directed along the phenol-benzimidazole bend mode.} 
  \label{fig:si_e1pt_da_dist}
\end{figure}

\begin{figure}
  \includegraphics[width=2.8in]{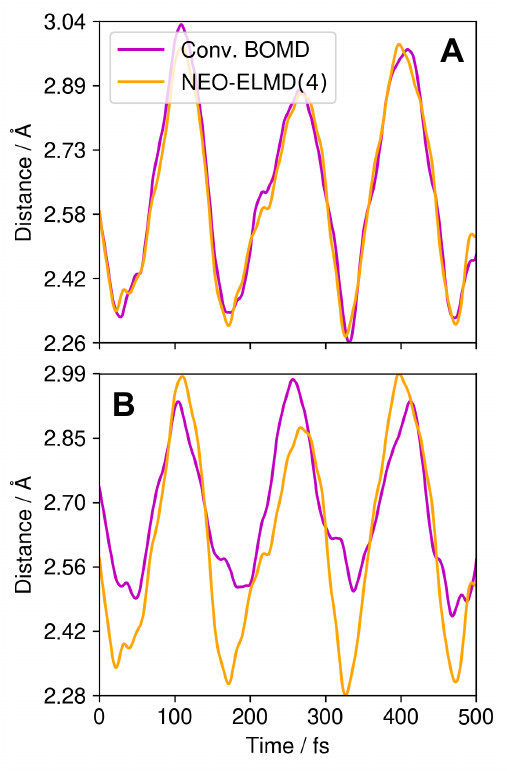} 
  \caption{\justifying Distances between the respective proton donors and acceptors along the BOMD (magenta) and NEO-ELMD$(4)$ (orange) E2PT trajectories for the (A) phenolic and (B) imidazolic proton transfers.}
  \label{fig:si_e2pt_da_dist}
\end{figure}

\begin{figure}
  \includegraphics[width=2.8in]{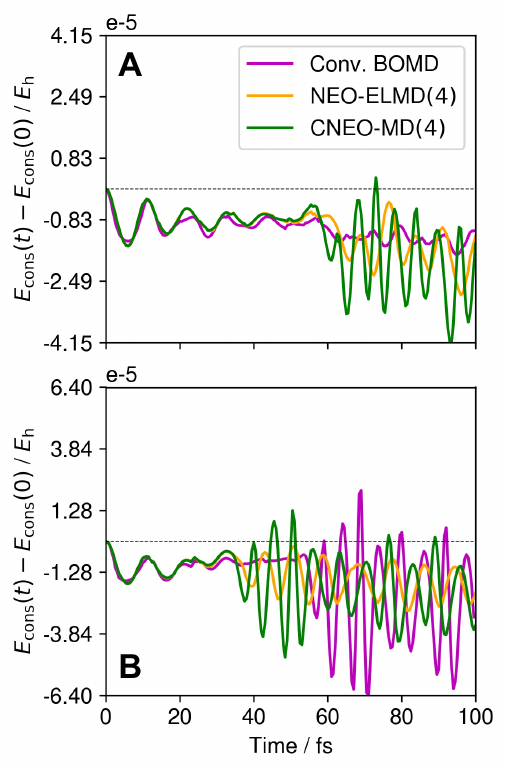} 
  \caption{\justifying Change in the conserved energy along the conventional BOMD (magenta), NEO-ELMD$(4)$ (orange), and CNEO-MD$(4)$ (green) E1PT trajectories, initialized with (A) zero velocities and (B) velocities along the phenol-benzimidazole bend mode.}
  \label{fig:si_e1pt_eng_cons}
\end{figure}

\begin{figure}
  \includegraphics[width=2.8in]{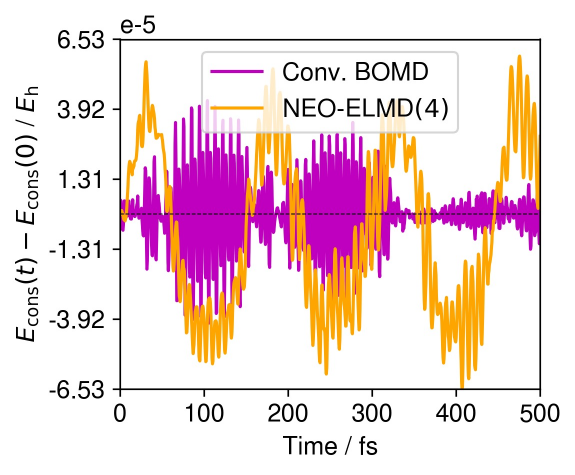} 
  \caption{\justifying Change in the conserved energy along the conventional BOMD (magenta) and NEO-ELMD$(4)$ (orange) E2PT trajectories.}
  \label{fig:si_e2pt_eng_cons}
\end{figure}

\begin{figure}
  \includegraphics[width=2.8in]{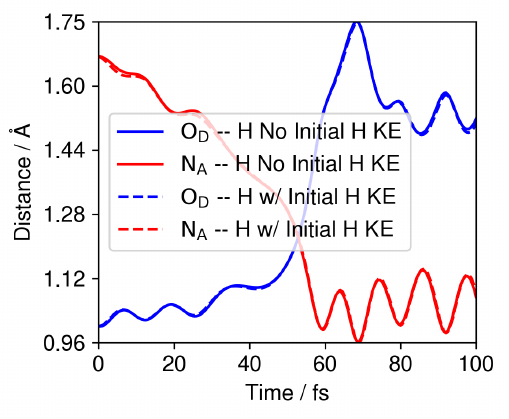} 
  \caption{\justifying Distance between the oxygen donor (blue) or nitrogen acceptor (red) and the classical proton along the E1PT conventional BOMD trajectory, where the velocity of the classical proton was set to zero (solid lines) or left unchanged (dashed lines) after velocity rescaling. The solid and dashed lines are virtually indistinguishable.}
  \label{fig:si_e1pt_pvel_v_no_pvel}
\end{figure}

\begin{figure}
  \includegraphics[width=2.8in]{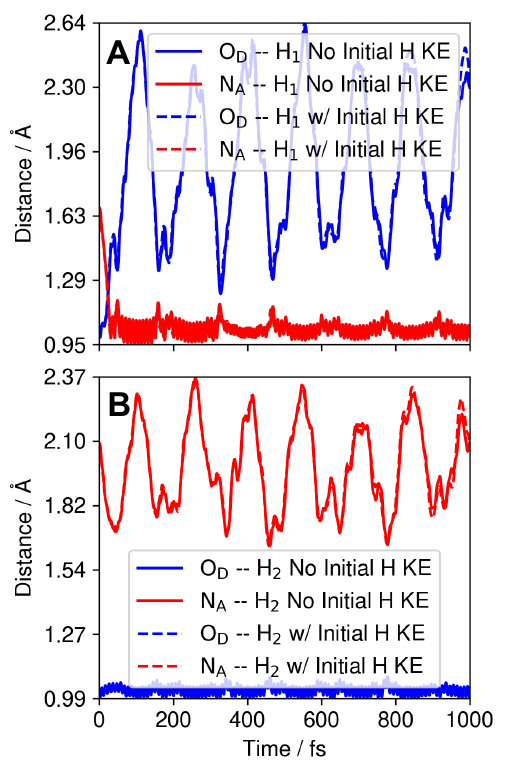} 
  \caption{\justifying Distance between the respective donors (blue) or acceptors (red) and the classical (A) phenolic and (b) imidazolic proton along the E2PT conventional BOMD trajectory, where the velocities of the classical protons were set to zero (solid lines) or left unchanged (dashed lines) after velocity rescaling. The solid and dashed lines are virtually indistinguishable.}
  \label{fig:si_e2pt_pvel_v_no_pvel}
\end{figure}

In this subsection, we provide additional information regarding the BIP trajectories analyzed in the main text. Fig.~\ref{fig:si_conv_bip_trajs} shows select structures along the conventional BOMD trajectories for the E1PT and E2PT BIP systems. Figs.~\ref{fig:si_e1pt_da_dist} and~\ref{fig:si_e2pt_da_dist} show the distances between the proton donors and acceptors for the E1PT and E2PT trajectories, respectively. Figs.~\ref{fig:si_e1pt_eng_cons} and~\ref{fig:si_e2pt_eng_cons} show the energy conservation analysis for the E1PT and E2PT trajectories, respectively. For these plots, the conserved energy is the physical energy in conventional BOMD and the extended energy for the NEO trajectories. Figs.~\ref{fig:si_e1pt_pvel_v_no_pvel} and~\ref{fig:si_e2pt_pvel_v_no_pvel} show the results of the conventional BOMD trajectories for the E1PT and E2PT systems, respectively, where the initial velocities of the transferring proton(s) were not set to zero after rescaling the velocities to the target temperature, as they were for the results presented in the main text. These plots demonstrate that setting the initial velocities of the transferring proton(s) to zero had a negligible impact on the dynamics, presumably because the  dynamics were dominated by heavy-atom motion.

\FloatBarrier

\FloatBarrier
\clearpage
\section{Timings of Select Trajectories}\label{sec:si_timings}
\FloatBarrier

In this section, we provide timings of NEO-BOMD, NEO-ELMD$(K)$, and CNEO-MD$(K)$ trajectories propagated with varying values of $K$. 

\begin{table}[H]
\begin{threeparttable}
\caption{\justifying CPU times\tnote{a} of NEO-BOMD trajectories with $\Delta t = 0.5$ fs and convergence criteria of $10^{-6}$ $E_{\mr{h}}$, propagated for 60 fs.\tnote{b}}
\renewcommand{\arraystretch}{1.5}
\begin{ruledtabular}
\begin{tabular}{c @{\hspace{12pt}}|@{\hspace{12pt}} c c c c}\label{tab:si_bomd_cpu_times}
Gradient Tolerance / $E_{\mathrm{h}}a_0^{-1}$ & PB4-F2 & PB5-G & PB6-H & 8s8p8d8f8g \\
\hline
$3.0\times10^{-4}$ & 98.1570 & 152.4907 & 343.5200 & 595.7902 \\
$1.2\times10^{-4}$ & 170.3615 & 163.0287 & 522.9248 & 957.2174 \\
$3.0\times10^{-5}$ & 255.3828 & 266.5062 & 880.2709 & 1637.9452 \\
\end{tabular}
\end{ruledtabular}
\begin{tablenotes}
\item[a] \footnotesize All CPU times are given in hours.
\item[b] \footnotesize Calculations performed on Intel Xeon Platinum 8268 (Cascade Lake) CPU node.
\end{tablenotes}
\end{threeparttable}
\end{table}

Table~\ref{tab:si_bomd_cpu_times} shows the CPU times of NEO-BOMD trajectories across protonic basis sets and gradient tolerance values with a time step of 0.5 fs and convergence criteria of $10^{-6}$ $E_{\mr{h}}$. As shown in Fig.~\ref{fig:si_bomd_eng_cons} and discussed in Section~\ref{subsec:si_more_trajs_malon}, the tightest level of gradient tolerance for the protonic basis function center optimization is necessary for sufficient energy conservation in NEO-BOMD, which is usually about two to three times more expensive than calculations that use the Q-Chem default value for gradient tolerance. Typically, as the protonic basis set becomes larger for a given tolerance, the computational expense increases. However, this trend may not always be followed, as the topologies of the extended NEO vibronic surfaces can vary between basis sets, making it possible for optimizations on the extended vibronic surface with smaller protonic basis sets to be more difficult than those with larger protonic basis sets. In some cases, smaller protonic basis sets result in more local minima along the dimensions corresponding to the protonic basis function center positions $\mc{R}_I$, explaining why the PB4-F2 and PB5-G trajectories do not have very different timings. However, by the time the protonic basis sets reach the size of PB6-H, the expense of each SCF procedure begins to dominate, making the PB6-H trajectories much more expensive than the PB5-G trajectories. These trends are not general but explain the observed timings for these specific cases. 

\begin{table}[H]
\begin{threeparttable}
\caption{\justifying CPU times\tnote{a} of NEO-ELMD$(K)$ and CNEO-MD$(K)$ trajectories with $\Delta t = 0.5$ fs and convergence criteria of $10^{-6}$ $E_{\mr{h}}$, propagated for 60 fs.\tnote{b}}
\renewcommand{\arraystretch}{1.5}
\begin{ruledtabular}
\begin{tabular}{c @{\hspace{12pt}}|@{\hspace{12pt}} c c c c @{\hspace{12pt}}|@{\hspace{12pt}} c c c c}\label{tab:si_abomd_cpu_times}
& \multicolumn{4}{c @{\hspace{12pt}}|@{\hspace{12pt}}}{NEO-ELMD$(K)$}
& \multicolumn{4}{c}{CNEO-MD$(K)$} \\
$K$ & PB4-F2 & PB5-G & PB6-H & 8s8p8d8f8g & PB4-F2 & PB5-G & PB6-H & 8s8p8d8f8g \\
\hline
0 & N/A\tnote{c} & N/A\tnote{c} & 4.5739 & 10.8954 & 1.6689 & 2.3819 & 3.9847 & 9.8837 \\
2 & 1.7464 & 2.2875 & 3.8430 & 9.2807 & 1.2598 & 1.8689 & 3.4647 & 8.6340 \\
4 & 1.6231 & 2.1122 & 3.6632 & 8.8949 & 1.1710 & 1.8210 & 3.2809 & 8.5716 \\
6 & 1.5524 & 2.1121 & 3.6023 & 8.8105 & 1.2378 & 1.7853 & 3.3284 & 8.7849 \\
\end{tabular}
\end{ruledtabular}
\begin{tablenotes}
\item[a] \footnotesize All CPU times are given in hours.
\item[b] \footnotesize Calculations performed on Intel Xeon Platinum 8562Y+ (Emerald Rapids) CPU node.
\item[c] \footnotesize Trajectory did not complete due to failed NEO-SCF.
\end{tablenotes}
\end{threeparttable}
\end{table}

Table~\ref{tab:si_abomd_cpu_times} provides the CPU times of the NEO-ELMD$(K)$ and CNEO-MD$(K)$ trajectories propagated with varying values of $K$ using a time step of 0.5 fs and convergence criteria of $10^{-6}$ $E_{\mr{h}}$. These timings show that the density matrix extrapolation procedure cuts down the total simulation time compared to $K=0$, where $K=4$ and $K=6$ both seem to outperform $K=2$ and offer similar speedups. Note that the total number of SCF iterations at each time step across all the trajectories utilizing density matrix extrapolation is lower than the number of iterations for the trajectories that do not use extrapolation, but the CPU times are not necessarily reflective of this aspect because the main computational bottleneck in these malonaldehyde trajectories is the analytical gradient calculation, rather than the SCF procedure. Also, note that for some trajectories, higher orders of extrapolation may not always lead to a decrease in simulation time. We observe this behavior for the CNEO-MD$(4)$ to CNEO-MD$(6)$ timings. Although the computational overhead of increasing $K$ (i.e., storing increasingly more previous density matrices) is quite low, the extrapolation is not always guaranteed to yield better initial guesses for the next time step with increasing $K$. We find $K=4$ to be sufficient in the speedup it provides while maintaining the fidelity of the initial guess.

\begin{table}[H]
\begin{threeparttable}
\caption{\justifying CPU times of NEO-ELMD$(K)$ trajectories for the BIP simulations initialized with non-zero velocities, propagated for 50 fs.\tnote{a}}
\renewcommand{\arraystretch}{1.5}
\begin{ruledtabular}
\begin{tabular}{c @{\hspace{12pt}}|@{\hspace{12pt}} c c @{\hspace{12pt}}|@{\hspace{12pt}} c c}\label{tab:si_bip_cpu_times}
& \multicolumn{2}{c @{\hspace{12pt}}|@{\hspace{12pt}}}{E1PT}
& \multicolumn{2}{c}{E2PT} \\
$K$ & CPU Time / hrs & Speedup Rel. to $K=0$ / \% & CPU Time / hrs & Speedup Rel. to $K=0$ / \%  \\
\hline
0 & 156.87 & - & 469.80 & - \\
2 & 137.54 & 12.32 & 471.06 & -0.26  \\
4 & 135.76 & 13.45 & 300.62 & 36.01  \\
\end{tabular}
\end{ruledtabular}
\begin{tablenotes}
\item[a] \footnotesize Calculations performed on Intel Xeon Platinum 8268 (Cascade Lake) CPU node.
\end{tablenotes}
\end{threeparttable}
\end{table}

Table~\ref{tab:si_bip_cpu_times} provides the CPU times of NEO-ELMD$(K)$ trajectories for the BIP simulations initialized with non-zero velocities propagated with varying values of $K$ using a time step of 0.5 fs and convergence criteria of $10^{-7}$ $E_{\mr{h}}$. To quickly assess the relative speedups compared to the $K=0$ trajectories, we only propagated these trajectories for 50 fs. For the E1PT system, we observe modest speedups on the order of what we observe in the malonaldehyde trajectories $(\approx12-16\%)$, where the speedup monotonically increases with $K$. However, for the E2PT system, we observe a very small slowdown with $K=2$, but a fairly large speedup of $\approx36\%$ with $K=4$. These E2PT results show how any given order of extrapolation is not guaranteed to result in a speedup, but higher orders of extrapolation most likely will result in a speedup, particularly for systems where the NEO-SCF procedure involves multiple quantum protons, as is the case for the E2PT system.

\section{Cartesian Coordinates}\label{sec:si_cart_coord}

\noindent In this section, we provide relevant Cartesian coordinates of the systems studied in this work.

\bigskip

\begingroup
\renewcommand{\arraystretch}{1.5}

\centering
\begin{threeparttable}
\captionof{table}{\justifying Optimized structure of malonaldehyde at the $\omega$B97X/def2-SVP level of theory.\tnote{a}}
\begin{ruledtabular}
\begin{tabular}{c c c c}\label{tab:si_malon_equil}
Atom & X & Y & Z \\
\hline
H & -0.6830093635 & \phantom{-}4.8873936970 & \phantom{-}0.0000000000 \\
C & -2.2965024834 & \phantom{-}1.7761906527 & \phantom{-}0.0000000000 \\
C & -0.0778296218 & \phantom{-}0.4606600999 & \phantom{-}0.0000000000 \\
C & \phantom{-}2.2683699385 & \phantom{-}1.8502803018 & \phantom{-}0.0000000000 \\
O & -2.4548505732 & \phantom{-}4.2439921132 & \phantom{-}0.0000000000 \\
O & \phantom{-}2.3773400124 & \phantom{-}4.1728718529 & \phantom{-}0.0000000000 \\
H & -4.1386563233 & \phantom{-}0.8241552831 & \phantom{-}0.0000000000 \\
H & -0.0809127358 & -1.5993342589 & \phantom{-}0.0000000000 \\
H & \phantom{-}4.0571330701 & \phantom{-}0.7432112784 & \phantom{-}0.0000000000 \\
\end{tabular}
\end{ruledtabular}
\begin{tablenotes}
\item[a] \footnotesize All values are in $a_0$.
\end{tablenotes}
\end{threeparttable}

\bigskip
\bigskip

\centering
\begin{threeparttable}
\captionof{table}{\justifying Transition state structure of malonaldehyde at the $\omega$B97X/def2-SVP level of theory.\tnote{a}}
\begin{ruledtabular}
\begin{tabular}{c c c c}\label{tab:si_malon_ts}
Atom & X & Y & Z \\
\hline
H & -0.0178400642 & \phantom{-}4.6948974682 & \phantom{-}0.0000000000 \\
C & -2.2738492601 & \phantom{-}1.8442730308 & \phantom{-}0.0000000000 \\
C & -0.0444835260 & \phantom{-}0.4162240616 & \phantom{-}0.0000000000 \\
C & \phantom{-}2.2034933679 & \phantom{-}1.8155603217 & \phantom{-}0.0000000000 \\
O & -2.2455272415 & \phantom{-}4.2398871387 & \phantom{-}0.0000000000 \\
O & \phantom{-}2.2062821788 & \phantom{-}4.2109971182 & \phantom{-}0.0000000000 \\
H & -4.1427804002 & \phantom{-}0.9160563337 & \phantom{-}0.0000000000 \\
H & -0.0583443778 & -1.6419483513 & \phantom{-}0.0000000000 \\
H & \phantom{-}4.0603662493 & \phantom{-}0.8634808204 & \phantom{-}0.0000000000 \\
\end{tabular}
\end{ruledtabular}
\begin{tablenotes}
\item[a] \footnotesize All values are in $a_0$.
\end{tablenotes}
\end{threeparttable}

\endgroup

\begin{table*}
\begin{threeparttable}
\caption{\justifying Optimized structure of neutral singlet E1PT BIP at the B3LYP-D3(BJ)/6-31G** level of theory.\tnote{a}}
\renewcommand{\arraystretch}{0.9}
\begin{ruledtabular}
\begin{tabular}{c c c c}\label{tab:si_e1pt}
Atom & X & Y & Z \\
\hline
H & \phantom{-}\phantom{-}1.9998484428 & \phantom{-}4.0074680603 & -0.0075081175 \\
C & \phantom{-}\phantom{-}7.0112302694 & -1.4835836654 & \phantom{-}0.0209628768 \\
C & \phantom{-}\phantom{-}6.8885244091 & \phantom{-}1.1896865648 & \phantom{-}0.0146639266 \\
C & \phantom{-}\phantom{-}9.1135165656 & \phantom{-}2.6145188763 & \phantom{-}0.0166050021 \\
C & \phantom{-}11.4019849190 & \phantom{-}1.3186321456 & \phantom{-}0.0248282327 \\
C & \phantom{-}11.4930869998 & -1.3395785949 & \phantom{-}0.0311771395 \\
C & \phantom{-}\phantom{-}9.2961700214 & -2.7916419113 & \phantom{-}0.0293547479 \\
C & \phantom{-}\phantom{-}2.9993056547 & -0.1517697733 & \phantom{-}0.0077468992 \\
H & \phantom{-}\phantom{-}9.0340853464 & \phantom{-}4.6623743272 & \phantom{-}0.0116495831 \\
H & \phantom{-}13.1589068709 & \phantom{-}2.3765445139 & \phantom{-}0.0263728194 \\
H & \phantom{-}13.3146890634 & -2.2816593868 & \phantom{-}0.0375746303 \\
N & \phantom{-}\phantom{-}4.3819780033 & \phantom{-}1.9438818218 & \phantom{-}0.0067461421 \\
C & \phantom{-}-1.0925415715 & \phantom{-}2.1296693142 & -0.0123011933 \\
C & \phantom{-}\phantom{-}0.2518131497 & -0.1939610505 & -0.0005739643 \\
C & \phantom{-}-1.0827553285 & -2.4835142124 & \phantom{-}0.0016670131 \\
C & \phantom{-}-3.7052251173 & -2.5605701008 & -0.0086751495 \\
C & \phantom{-}-4.9752141226 & -0.2316020140 & -0.0200950093 \\
C & \phantom{-}-3.7671449147 & \phantom{-}2.1131765625 & -0.0221870952 \\
H & \phantom{-}-0.0395022924 & -4.2498426050 & \phantom{-}0.0117648626 \\
H & \phantom{-}-7.0146141811 & -0.2487190426 & -0.0276948756 \\
C & \phantom{-}-5.0825915909 & -5.1135036694 & -0.0051976261 \\
C & \phantom{-}-5.2810935605 & \phantom{-}4.5948138133 & -0.0346543204 \\
O & \phantom{-}0.1422622254 & \phantom{-}4.3537808070 & -0.0146179565 \\
N & \phantom{-}4.5140056497 & -2.2715382158 & \phantom{-}0.0165012929 \\
H & \phantom{-}3.9058963212 & -4.0730837876 & \phantom{-}0.0178150752 \\
C & \phantom{-}-4.3079108868 & -6.6378157302 & -2.3691383971 \\
H & \phantom{-}-2.2725615316 & -6.9985461832 & -2.4142787938 \\
H & \phantom{-}-5.2709289801 & -8.4699819444 & -2.3983577639 \\
H & \phantom{-}-4.8048235471 & -5.6159543869 & -4.0970379635 \\
C & \phantom{-}-4.3500047786 & -6.6071277397 & \phantom{-}2.3913512349 \\
H & \phantom{-}-4.8792885925 & -5.5635719352 & \phantom{-}4.0965541756 \\
H & \phantom{-}-5.3117611001 & -8.4399368560 & \phantom{-}2.4266707102 \\
H & \phantom{-}-2.3153425667 & -6.9640998596 & \phantom{-}2.4776832171 \\
C & \phantom{-}-7.9715712283 & -4.8037765892 & -0.0334797204 \\
H & \phantom{-}-8.6463269513 & -3.7788305522 & \phantom{-}1.6311535265 \\
H & \phantom{-}-8.6143966140 & -3.7950057477 & -1.7205242566 \\
H & \phantom{-}-8.8766067908 & -6.6637207255 & -0.0328473570 \\
C & \phantom{-}-4.6345369486 & \phantom{-}6.1374410750 & -2.4292829666 \\
H & \phantom{-}-5.7095986847 & \phantom{-}7.9067559522 & -2.4345102584 \\
H & \phantom{-}-2.6275812565 & \phantom{-}6.5897318587 & -2.5108324876 \\
H & \phantom{-}-5.1411745615 & \phantom{-}5.0757416096 & -4.1321117344 \\
C & \phantom{-}-4.6527632092 & \phantom{-}6.1497010384 & \phantom{-}2.3569226204 \\
H & \phantom{-}-2.6470972023 & \phantom{-}6.6057246946 & \phantom{-}2.4493381998 \\
H & \phantom{-}-5.7311456855 & \phantom{-}7.9169389292 & \phantom{-}2.3470165925 \\
H & \phantom{-}-5.1685186208 & \phantom{-}5.0947760496 & \phantom{-}4.0612105807 \\
C & \phantom{-}-8.1498666383 & \phantom{-}4.1190343999 & -0.0444954562 \\
H & \phantom{-}-8.7578632780 & \phantom{-}3.0718532795 & -1.7222449475 \\
H & \phantom{-}-8.7712837052 & \phantom{-}3.0825575749 & \phantom{-}1.6349975324 \\
H & \phantom{-}-9.1380560316 & \phantom{-}5.9354170786 & -0.0541450736 \\
H & \phantom{-}\phantom{-}9.3711665911 & -4.8411480747 & \phantom{-}0.0341681182 \\
\end{tabular}
\end{ruledtabular}
\begin{tablenotes}
\item[a] \footnotesize All values are in $a_0$.
\end{tablenotes}
\end{threeparttable}
\end{table*}

\begin{table*}
\begin{threeparttable}
\caption{\justifying Optimized structure of neutral singlet E2PT BIP at the B3LYP-D3(BJ)/6-31G** level of theory.\tnote{a}}
\renewcommand{\arraystretch}{0.3}
\begin{ruledtabular}
\begin{tabular}{c c c c}\label{tab:si_e2pt}
Atom & X & Y & Z \\
\hline
H & \phantom{-}-2.0832318054 & \phantom{-}5.1852941954 & -0.0274308340 \\
H & \phantom{-}\phantom{-}3.2162524054 & -1.3375434781 & \phantom{-}0.0000181365 \\
C & \phantom{-}\phantom{-}4.7529572358 & \phantom{-}2.2819186769 & -0.0100833117 \\
C & \phantom{-}\phantom{-}3.5389809525 & \phantom{-}4.6685489676 & -0.0208525434 \\
C & \phantom{-}\phantom{-}4.9741255762 & \phantom{-}6.8820838309 & -0.0255705860 \\
C & \phantom{-}\phantom{-}7.5961249456 & \phantom{-}6.6297413533 & -0.0189545761 \\
C & \phantom{-}\phantom{-}8.7714722106 & \phantom{-}4.2548391527 & -0.0077080445 \\
C & \phantom{-}\phantom{-}7.3923646466 & \phantom{-}1.9870802148 & -0.0026797328 \\
C & \phantom{-}\phantom{-}0.5717386353 & \phantom{-}1.8230188803 & -0.0166950475 \\
H & \phantom{-}\phantom{-}4.0675740958 & \phantom{-}8.7196888346 & -0.0339833865 \\
H & \phantom{-}\phantom{-}8.7706229665 & \phantom{-}8.3106193745 & -0.0224218219 \\
H & \phantom{-}10.8187353558 & \phantom{-}4.1955527293 & -0.0029862599 \\
N & \phantom{-}\phantom{-}0.9484454340 & \phantom{-}4.3128411515 & -0.0247217234 \\
C & \phantom{-}-4.0995799122 & \phantom{-}2.1793185549 & -0.0212544562 \\
C & \phantom{-}-1.9047029078 & \phantom{-}0.6364998326 & -0.0165603638 \\
C & \phantom{-}-2.1479885483 & -2.0012625347 & -0.0108926693 \\
C & \phantom{-}-4.4971752382 & -3.1708963385 & -0.0093606131 \\
C & \phantom{-}-6.6304530767 & -1.5941502978 & -0.0138480005 \\
C & \phantom{-}-6.5196712534 & \phantom{-}1.0411397123 & -0.0198204408 \\
H & \phantom{-}-0.4510202452 & -3.1524620882 & -0.0081176842 \\
H & \phantom{-}-8.4727358276 & -2.4669974279 & -0.0126201660 \\
C & \phantom{-}-4.6773804991 & -6.0672968103 & -0.0030408169 \\
C & \phantom{-}-8.9297626659 & \phantom{-}2.6674456131 & -0.0241189486 \\
O & \phantom{-}-3.9140847815 & \phantom{-}4.7169020657 & -0.0266200778 \\
N & \phantom{-}\phantom{-}2.8262087942 & \phantom{-}0.5360226309 & -0.0080892986 \\
C & \phantom{-}-3.3575018446 & -7.1260910245 & -2.3796306556 \\
H & \phantom{-}-1.3566452482 & -6.6096861633 & -2.4345308992 \\
H & \phantom{-}-3.4771433360 & -9.1929055608 & -2.4152588341 \\
H & \phantom{-}-4.2447822250 & -6.3956116019 & -4.0987401606 \\
C & \phantom{-}-3.3571596397 & -7.1158303298 & \phantom{-}2.3779077075 \\
H & \phantom{-}-4.2433624277 & -6.3770227616 & \phantom{-}4.0940147167 \\
H & \phantom{-}-3.4779164625 & -9.1824121841 & \phantom{-}2.4231467064 \\
H & \phantom{-}-1.3559831722 & -6.6003133985 & \phantom{-}2.4296938489 \\
C & \phantom{-}-7.4295897870 & -6.9997200382 & -0.0009006942 \\
H & \phantom{-}-8.4564989823 & -6.3504121015 & \phantom{-}1.6728768926 \\
H & \phantom{-}-8.4564218390 & -6.3580384876 & -1.6776525033 \\
H & \phantom{-}-7.4662332084 & -9.0678548390 & \phantom{-}0.0038182738 \\
C & \phantom{-}-8.9956943314 & \phantom{-}4.3335995745 & -2.4214097849 \\
H & -10.7214458817 & \phantom{-}5.4776859083 & -2.4374868380 \\
H & \phantom{-}-7.3712486331 & \phantom{-}5.5960579687 & -2.5084838195 \\
H & \phantom{-}-8.9986350238 & \phantom{-}3.1518869097 & -4.1206004015 \\
C & \phantom{-}-8.9985192380 & \phantom{-}4.3412789164 & \phantom{-}2.3677440187 \\
H & \phantom{-}-7.3740769802 & \phantom{-}5.6039115247 & \phantom{-}2.4527310223 \\
H & -10.7242493542 & \phantom{-}5.4854530131 & \phantom{-}2.3780887435 \\
H & \phantom{-}-9.0035235058 & \phantom{-}3.1650271651 & \phantom{-}4.0707337678 \\
C & -11.3395580430 & \phantom{-}1.0392073734 & -0.0229142166 \\
H & -11.4633369668 & -0.1667366929 & -1.6992605478 \\
H & -11.4658199397 & -0.1603452963 & \phantom{-}1.6578132498 \\
H & -12.9949669090 & \phantom{-}2.2783386219 & -0.0264077644 \\
C & \phantom{-}\phantom{-}8.5656013335 & -0.5439986107 & \phantom{-}0.0101261727 \\
C & \phantom{-}11.1990517111 & -0.8994430911 & \phantom{-}0.0209164285 \\
C & \phantom{-}12.1782326326 & -3.3331549918 & \phantom{-}0.0328873450 \\
H & \phantom{-}12.4632614759 & \phantom{-}0.7089962195 & \phantom{-}0.0202218418 \\
C & \phantom{-}\phantom{-}7.9461031921 & -4.8795579617 & \phantom{-}0.0231928344 \\
C & \phantom{-}10.5260509922 & -5.3875225844 & \phantom{-}0.0340403780 \\
H & \phantom{-}14.2090817780 & -3.6241503644 & \phantom{-}0.0413702567 \\
H & \phantom{-}\phantom{-}6.5825175112 & -6.4182405391 & \phantom{-}0.0238063051 \\
H & \phantom{-}11.2095194381 & -7.3191292003 & \phantom{-}0.0430916754 \\
N & \phantom{-}\phantom{-}6.9763464596 & -2.5479942534 & \phantom{-}0.0116275598 \\
\end{tabular}
\end{ruledtabular}
\begin{tablenotes}
\item[a] \footnotesize All values are in $a_0$.
\end{tablenotes}
\end{threeparttable}
\end{table*}

\end{document}